\documentclass[11pt]{article}

\usepackage[showframe=false,bottom=2.4cm, top=2.54cm, left=2.2cm, right=2.2cm]{geometry}

\usepackage[utf8]{inputenc}
\usepackage[T1]{fontenc}
\usepackage{amsfonts}
\usepackage{amsmath}
\usepackage{hyperref}
\usepackage[capitalise]{cleveref}
\usepackage{placeins}
\usepackage[labelfont=bf,font={small}]{caption}

\usepackage{cite} %
\newcommand{\citep}{\cite} %
\usepackage{graphicx}
\usepackage{booktabs} 
\usepackage{adjustbox}
\usepackage{subcaption}
\usepackage{setspace}

\usepackage{contour}
\usepackage{ulem}
\contourlength{0.8pt}
\newcommand{\myunderline}[1]{%
  \uline{\phantom{#1}}%
  \llap{\contour{white}{#1}}%
}

\usepackage{authblk}

\author[1]{Justin Engelmann\thanks{\texttt{justin.engelmann@ed.ac.uk}}}
\author[2]{Alice D. McTrusty}
\author[2]{Ian J. C. MacCormick}
\author[2]{Emma Pead}
\author[3]{\\Amos Storkey\thanks{Equally contributing senior authors.}}
\author[4]{Miguel O. Bernabeu\protect\footnotemark[2]}

\affil[1]{UKRI CDT Biomedical AI, University of Edinburgh}
\affil[2]{Centre for Clinical Brain Sciences, University of Edinburgh}
\affil[3]{Institute for Adaptive and Neural Computation, University of Edinburgh}
\affil[4]{Centre for Medical Informatics, University of Edinburgh}

\title{Detection of multiple retinal diseases in ultra-widefield fundus images using deep learning: data-driven identification of relevant regions}

\date{\vspace{-1em}}

\begin{document}
\maketitle

\begin{abstract}

Ultra-widefield (UWF) imaging is a promising modality that captures a larger retinal field of view compared to traditional fundus photography. Previous studies showed that deep learning (DL) models are effective for detecting retinal disease in UWF images, but primarily considered individual diseases under less-than-realistic conditions (excluding images with other diseases, artefacts, comorbidities, or borderline cases; and balancing healthy and diseased images) and did not systematically investigate which regions of the UWF images are relevant for disease detection. We first improve on the state of the field by proposing a DL model that can recognise multiple retinal diseases under more realistic conditions. We then use global explainability methods to identify which regions of the UWF images the model generally attends to. Our model performs very well, separating between healthy and diseased retinas with an area under the curve (AUC) of 0.9206 on an internal test set, and an AUC of 0.9841 on a challenging, external test set. When diagnosing specific diseases, the model attends to regions where we would expect those diseases to occur. We further identify the posterior pole as the most important region in a purely data-driven fashion. Surprisingly, 10\% of the image around the posterior pole is sufficient for achieving comparable performance to having the full images available.

\end{abstract}

\section{Introduction}

Retinal diseases are a significant public health burden. The associated loss of vision reduces the quality of life of affected patients \citep{brown1999vision} and has major economic impact \citep{pezzullo2018economic}. Age, lifestyle factors and some diseases (e.g. diabetes) are key risk factors for retinal diseases such as age-related macular degeneration (AMD) and diabetic retinopathy (DR). This burden is thus set to increase in ageing societies and in those that are rapidly urbanising and experience rising incidence of non-communicable diseases.
Retinal diseases are detected and diagnosed through eye examinations which often include fundus imaging. Traditional colour fundus photography (CFP) typically images 30-60° of the retina, but new imaging devices allow for ultra-widefield (UWF) retinal images with a field of view of 100-200° \citep{doi:10.1177/2515841419899495, nagiel2016ultra}. Thus, UWF could capture retinal pathology that would be missed with CFP. This could enable more accurate screening and earlier detection. Sophisticated retinal imaging hardware is increasingly available in community optometry practice. Diagnosis and support systems are urgently needed to help practitioners make clinical decisions using these new imaging modalities, particularly for UWF retinal images that fewer clinicians are experienced with.

Fortunately, recent advances in deep learning (DL) make it feasible to automatically process images with similar performance to human experts on many tasks. Previous work showed that DL models are effective at detecting disease in UWF images but the evaluation approaches employed in literature to date have been criticised as unrealistic by clinical practitioners \citep{tan2020deep}. For example, models were typically trained to detect only a single specific disease and then evaluated on a dataset that contained equal numbers of images showing that disease and healthy controls, but no other diseases \citep{matsuba2019accuracy, nagasato2019deep, nagasato2018deep, tabuchi2018discrimination, masumoto2018deep, nagasawa2018accuracy, nagasawa2019accuracy, ohsugi2017accuracy,masumoto2018retinal,masumoto2019accuracy}. Recent work proposed a model for three diseases, but also excluded other diseases and artificially balanced the data \citep{Antakibjophthalmol-2021-319030}. Images with artefacts or multiple diseases were also excluded \citep{matsuba2019accuracy, nagasato2019deep, nagasato2018deep, tabuchi2018discrimination, masumoto2018deep, nagasawa2018accuracy, nagasawa2019accuracy, ohsugi2017accuracy,masumoto2018retinal,masumoto2019accuracy,Antakibjophthalmol-2021-319030}. We improve on this approach by proposing a model that can detect seven different retinal diseases and under more realistic conditions than what has previously been considered. Particularly, we consider several retinal diseases without excluding images with multiple diseases in the same image, artefacts or borderline cases. We further do not artificially balance the data by discarding healthy cases. 

Thus far, there has been no systematic investigation of which UWF image regions are relevant for DL model-based detection of retinal disease. Previous work compared the performance of DL models for detecting a single disease (glaucoma) using either full or optic disc cropped UWF images \citep{tabuchi2018discrimination} -- but in this case the optic disc was selected based on domain knowledge. Similar work has been undertaken for CFP but again the regions of interest were defined using domain knowledge \citep{hemelings2021deep}.We instead investigate which regions the DL model attends to for seven different diseases through a data-driven analysis. This serves both to understand how DL models use UWF images and to validate that the DL model works as intended rather than relying on undesirable shortcut artefacts. We further investigate how model performance degrades when removing either the least or the most important regions.

We find that the proposed DL model performs very well on an internal test set even when this is not artificially balanced and includes images with artefacts, borderline cases, and comorbidities. Our model also outperforms an ensemble of binary classifiers trained on balanced data for individual diseases which is the prevailing approach in the literature.  Evaluation on a challenging external test set which includes images with different preprocessing and images taken with a variety of UWF imaging devices also evidences very high real-world performance. For individual diseases, the model attends to regions that are consistent with domain knowledge, indicating that the model works as intended. Interestingly, we find that all seven diseases are detectable in the posterior pole and arrive at this conclusion in a purely data-driven way.

\section{Results}

\subsection{Study data characteristics and data split }
The Tsukazaki Optos Public (TOP) dataset consists of 13,047 UWF retina images of 8,588 eyes belonging to 5,389 patients that were taken between October 11, 2011 and September 6, 2018 at Tsukazaki hospital in Himeji, Japan.
All images were taken with an 
Optos{\textcopyright} 200Tx (Optos\textcopyright, Dunfermline, Scotland, U.K.) ultra-widefield scanning laser ophthalmoscopy imaging device. The data was released by Dr. Hiroki Masumoto for research use and this was approved by the Ethics Committee of Tsukazaki Hospital. Each image has binary labels for eight retinal diseases: Artery Occlusion (AO), Age-related Macular Degeneration (AMD), Diabetic Retinopathy (DR), Glaucoma (Gla), Macular Hole (MH), Retinal Detachment (RD), Retinitis Pigmentosa (RP), Retinal Vein Occlusion (RVO). For each image, the dataset also contains a unique patient ID, age in years, sex, whether the image is of a left or right eye, and the patient’s binary diabetes status. 50\% of the patients are female, and 50\% of the images are of left eyes. The average patient age is 65.83 years with a standard deviation of 12.99 years.
\cref{tab:datasplit_stratvalues_patientlevel} shows the numbers of patients per disease. Additional exploratory data analysis can be found in \cref{sec:sup_eda_TOP}. Notably, there are some signs of possible selection bias: half of the patients are female despite most patients being of advanced age where we would expect more females, older patients tend to be healthier, and patients with diabetes appear at lower risk for non-DR retinal disease. Prior work on the TOP dataset by ophthalmologists noted that the dataset contains both obvious and subtle cases, at least for the diseases that were considered there (RD, RP, RVO). \citep{Antakibjophthalmol-2021-319030}

\begin{table}[!t]%
\caption{Overview of the TOP dataset and the three subsets. We report the number and share of patients by stratification value (see \cref{sec:methods_datasplit_detailed}), the number of patients and images, as well as mean age and sex ratio for each set.}
\label{tab:datasplit_stratvalues_patientlevel}
\centering
\begin{adjustbox}{max width=\textwidth, max totalheight=1\textheight-2\baselineskip}
{\small

\begin{tabular}{rllll}
\toprule
\multicolumn{1}{l}{{}}                       & TOP Dataset   & Train Set     & Validation Set & Test Set      \\
\midrule
\multicolumn{1}{l}{Patients classified as:} &               &               &               &               \\ 
Healthy                                     & 2322 (43.2\%) & 1625 (43.2\%) & 348  (43.2\%) & 349  (43.2\%) \\
Gla                                         & 943  (17.5\%) & 660  (17.5\%) & 141  (17.5\%) & 142  (17.6\%) \\
DR                                          & 682  (12.7\%) & 477  (12.7\%) & 102  (12.7\%) & 103  (12.8\%) \\
RVO                                         & 438  (8.1\%)  & 307  (8.2\%)  & 65   (8.1\%)  & 66   (8.2\%)  \\
RD                                          & 417  (7.8\%)  & 292  (7.8\%)  & 63   (7.8\%)  & 62   (7.7\%)  \\
AMD                                         & 285  (5.3\%)  & 200  (5.3\%)  & 43   (5.3\%)  & 42   (5.2\%)  \\
MH                                          & 179  (3.3\%)  & 125  (3.3\%)  & 27   (3.3\%)  & 27   (3.3\%)  \\
RP                                          & 110  (2.0\%)  & 77   (2.0\%)  & 17   (2.1\%)  & 16   (2.0\%)  \\ 
\midrule
\multicolumn{1}{l}{Total Patients}          & 5376          & 3763          & 806           & 807           \\ 
\midrule
\multicolumn{1}{l}{Total Images}        & 13026         & 9121          & 1911          & 1994          \\
\multicolumn{1}{l}{Patient Age (Mean ± Std)} & 65.83 ± 12.99 & 65.72 ± 13.04 & 65.89 ± 12.84  & 66.29 ± 12.87 \\
\multicolumn{1}{l}{Female Patients}  & 2669 (50\%)   & 1843 (49\%)   & 392 (49\%)     & 434 (54\%)    \\
\bottomrule
\end{tabular}

}
\end{adjustbox}
\end{table}%
 
Apart from the 21 images showing AO, we did not exclude any images from the study. These images were excluded as 21 is too few to both train and meaningfully evaluate a model on that label.\footnote{Instead, they were used as an additional held-out test set to assess whether our model generalises to unseen diseases, see \cref{sec:sup_AO_results}.} Common reasons for excluding images in the literature are quality issues (e.g. reflection or eyelash artefacts, poor contrast), hard-to-recognise pathology (e.g. borderline cases), multiple diseases, or discarding of healthy controls to balance the labels \citep{matsuba2019accuracy, nagasato2019deep, nagasato2018deep, tabuchi2018discrimination, masumoto2018deep, nagasawa2018accuracy, nagasawa2019accuracy, ohsugi2017accuracy,masumoto2018retinal,masumoto2019accuracy, Antakibjophthalmol-2021-319030}. We decided against excluding images for any of those reasons as poor quality images, borderline cases, comorbidities, and label imbalance are clinical realities and - as recently highlighted by clinical practitioners \citep{tan2020deep} - DL models should be robust to those challenges. Furthermore, setting the bar for image quality high runs the risk of an overly optimistic estimate of DL model performance, whereas our approach is more likely to provide an estimate that is too pessimistic. We consider this preferable when developing methods intended for clinical applications.

We split the data into three sets - training, validation and test – containing 70, 15 and 15\% of the patients respectively. Including a validation set is an important methodological detail that has been omitted by some prior work on the TOP dataset \citep{matsuba2019accuracy, nagasato2019deep, nagasato2018deep, tabuchi2018discrimination, masumoto2018deep, nagasawa2018accuracy, nagasawa2019accuracy, ohsugi2017accuracy,masumoto2018retinal,masumoto2019accuracy}. Developing DL models is typically an iterative process and the validation set guides the process while keeping the test set unseen. Not using a validation set in machine learning is comparable to not controlling for multiple comparisons in statistics. We chose our final model on the validation set before ever making any test set evaluations.\footnote{The convention in the machine learning space is that the final held-out set used to estimate model performance is called ``test set'', whereas the set used to provide feedback for the modelling process is called ``validation set''. This clashes with terminology used in the medical community, where the evaluation on the test set is typically referred to as ``internal validation''. Note that the validation set (in the ML sense) is not used for internal validation (in the medical sense).} Furthermore, we split the data on the patient-level rather than image-level. Otherwise, different images of the same eye could end up in both the training and test sets, rewarding models that overfit. To ensure that the distributions of diseases are similar across all three sets, we split stratified by patient disease status (see \cref{sec:methods_datasplit_detailed}).

\subsection{The proposed DL model achieves state-of-the-art results on internal validation}

\begin{table}[!t]%
\caption{Test set performance (AUC) of baselines and final model for each label. AUC assesses how well a model can separate positive and negative samples for a given label. Images are weighted such that each eye contributes equally, even if a specific eye was imaged multiple times. Higher is better, best values in bold. }
\label{tab:test_aucs_eyelevel}
\centering
\begin{adjustbox}{max width=\textwidth, max totalheight=1\textheight-2\baselineskip}
{\small
\begin{tabular}{lllllllll}
\toprule
{} &         Diseased &               DR &              Gla &               RD &              RVO &              AMD &               RP &               MH \\
\midrule
Logistic Regression with Age + Sex &           0.5964 &           0.5988 &           0.5155 &           0.7676 &           0.4892 &           0.8021 &           0.6776 &           0.5625 \\
Ensemble of Experts (binary DL models + balanced data)       &           0.8318 * &           0.8432 &           0.9141 &           0.9217 &           0.8996 &           0.7113 &  \textbf{0.9490} &           0.6454 \\
Ours (Single multi-label DL model + realistic data)                           &  \textbf{0.9206} &  \textbf{0.9125} &  \textbf{0.9422} &  \textbf{0.9753} &  \textbf{0.9468} &  \textbf{0.9510} &           0.9438 &  \textbf{0.7987} \\
\bottomrule
\smallskip * Using maximum of individual predictions (\cref{sec:methods_baselines}). 
\end{tabular}

}
\end{adjustbox}
\end{table}%

\begin{table}[!t]%
\caption{Test set performance (Brier score) of baselines and final model for each label. Brier score is sensitive to how well a model's predicted probablities are calibrated. Images are weighted such that each eye contributes equally, even if a specific eye was imaged multiple times. Lower is better, best values in bold.}
\label{tab:test_brier_eyelevel}
\centering
\begin{adjustbox}{max width=\textwidth, max totalheight=1\textheight-2\baselineskip}
{\small
\begin{tabular}{lllllllll}
\toprule
{} &         Diseased &               DR &              Gla &               RD &              RVO &              AMD &               RP &               MH \\
\midrule
Logistic Regression with Age + Sex &           0.2522 &           0.1354 &           0.1580 &           0.0466 &           0.0567 &           0.0426 &           0.0242 &  \textbf{0.0227} \\
Ensemble of Experts (binary DL models + balanced data)       &           0.1919 * &           0.1421 &           0.1182 &           0.0780 &           0.1211 &           0.2086 &           0.0993 &           0.2373 \\
Ours (Single multi-label DL model + realistic data)                                &  \textbf{0.1144} &  \textbf{0.0690} &  \textbf{0.0645} &  \textbf{0.0150} &  \textbf{0.0283} &  \textbf{0.0269} &  \textbf{0.0081} &           0.0238 \\
\bottomrule
\smallskip * Using maximum of individual predictions (\cref{sec:methods_baselines}). 
\end{tabular}

}
\end{adjustbox}
\end{table}%

The DL model outputs a probability for each of the seven retinal diseases as well as a probability for the input retina being diseased generally. To allow our model to deal with images that show multiple diseases, we frame the problem as multi-label rather than multi-class classification. This allows the model to predict more than one disease with high confidence, if appropriate. Our approach is described in detail in \cref{sec:methods_model_problem_framing}.
We evaluated our model on the unseen internal test set. \cref{tab:test_aucs_eyelevel} shows the area under the receiver operating characteristic curve (AUC) for each label. We weigh images such that each eye contributes equally to the metrics, regardless of how many times it was imaged. Our model can discriminate between healthy and diseased retinas with an AUC of 0.9206 and achieves excellent separation for six diseases (AUCs 0.9125-0.9753) and good separation (AUC = 0.7987) for the seventh (MH). MH is the disease with the fewest images available and thus our model had the least examples to learn from. Furthermore, diagnosis of MH would commonly be confirmed using different types of imaging such as optical coherence tomography and can be hard to recognise from images alone \citep{duker2013international}.

We also consider two baselines (\cref{sec:methods_baselines}): First, a model using only the patient’s age and sex as variables but no image information. We include this as a simple baseline to ensure that the DL model does not only infer age and sex from the image to make its predictions. We considered RandomForest, k-Nearest Neighbours, and Logistic Regression as classification algorithms and found that Logistic Regression performed best on the validation set. Second, an “Ensemble of Experts” of binary models that are the same as our proposed model except that they each have a single output and were trained on balanced data containing only controls and a specific disease. This is conceptually akin to approaches that have been used in previous work \citep{matsuba2019accuracy, nagasato2019deep, nagasato2018deep, tabuchi2018discrimination, masumoto2018deep, nagasawa2018accuracy, nagasawa2019accuracy, ohsugi2017accuracy,masumoto2018retinal,masumoto2019accuracy}. We find that the DL models clearly outperform the Age+Sex baseline on all labels, except on AMD where the Ensemble of Experts performs the worst. The Ensemble of Experts achieves a slightly higher (+0.0052) but similar AUC on RP, but our proposed model substantially outperforms it on all other labels. We also assessed calibration through the Brier Score (\cref{tab:test_brier_eyelevel}) and find that the Ensemble of Experts is badly calibrated compared to our proposed model, with Brier scores on average being 5.6 times higher. This is a result of being trained on artificially balanced data.

\begin{figure}[!th]
     \centering
    \includegraphics[width=0.65\textwidth]{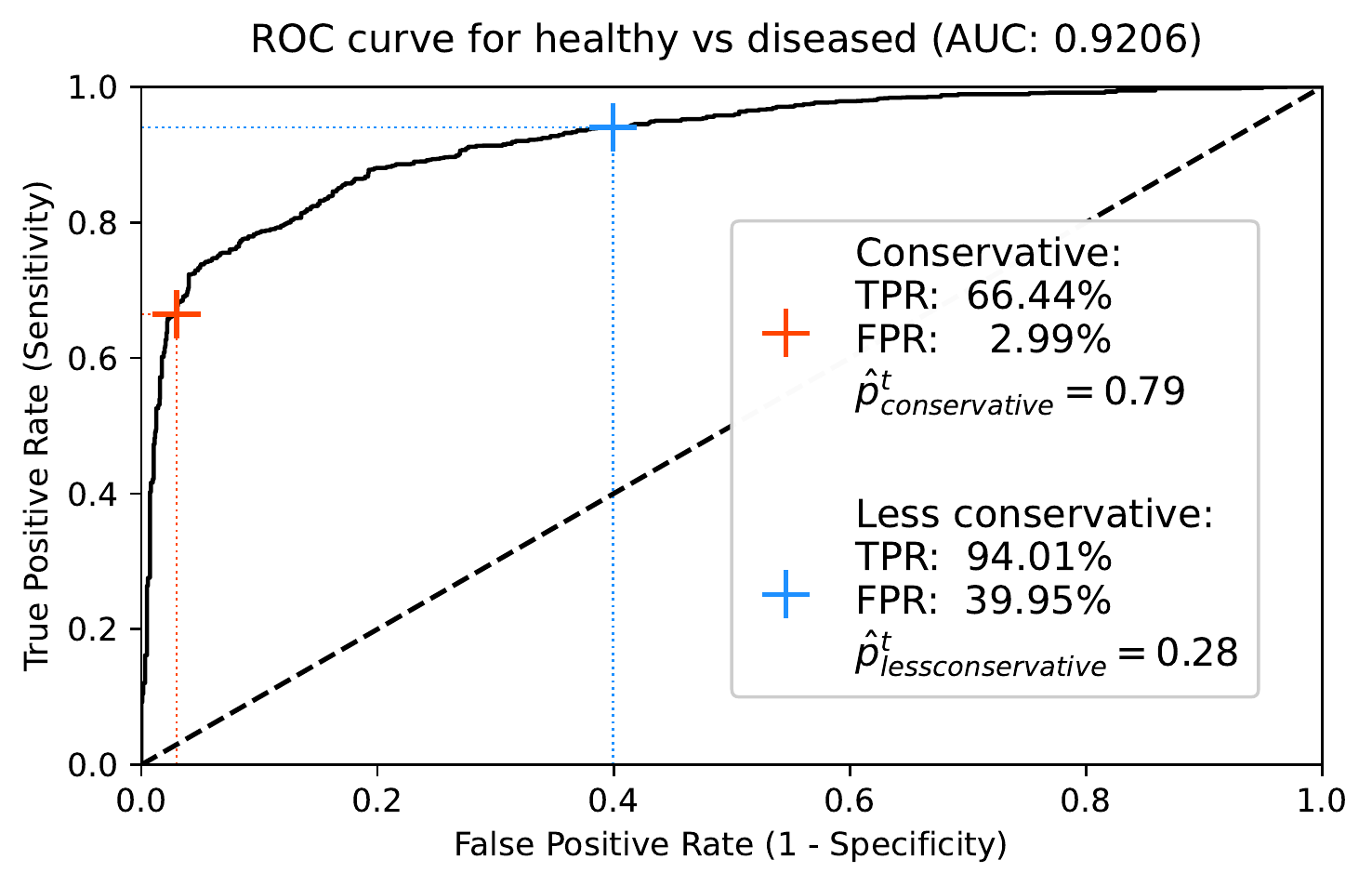}
\caption{ROC curve of our model predicting healthy vs diseased on the test set, weighted such that each eye contributes equally. Markers indicate where on the ROC curve we end up given the constraint of the respective use case. The dashed line is the identity line equivalent to random guessing.}
    \label{fig:ROC_usecases}

\end{figure}

In addition to technical measures of model performance, we also evaluated our model for two archetypical use cases to assess whether the model would be useful in practical applications \citep{tan2020deep}. First, a use case where false positives are costly and we thus want a conservative decision threshold $\hat{p}^t$ for flagging an image up as diseased. Second, a use case where false positives are less costly and we want a less conservative decision threshold. The conservative case could be an early screening application at a high street optician where a false positive might result in an unnecessary referral to an ophthalmologist which causes the patient stress and wastes healthcare resources. The less conservative case could be a clinical decision support system where an ophthalmologist can easily dismiss false positives and the focus is on providing a second opinion as well as reducing the chance of something being overlooked. In a concrete application, any point on the ROC curve can be chosen, and we could implement a complex, multi-level decision process. For example, we could use a traffic light-like system where a yellow alarm is raised for samples where $\hat{p}$ exceeds a less conservative threshold and a red alarm if it exceeds a conservative threshold. In an early screening context, the yellow alarm might then translate into simply scheduling the next routine scan sooner and while the red alarm means that the patient will be referred to an ophthalmologist. In any case, the clinician ultimately makes the diagnosis and decides on the next steps taking into account the available information. This includes their own assessment of the patient and the patient's health history and symptoms. A DL-based clinical decision support system would be an additional input to the clinician's decision making.

For the present work, however, our goal is to use the two archetypical use cases - a conservative and a less conservative case - to evaluate whether our model's performance at concrete decision thresholds would be potentially clinically useful. For the conservative and less conservative cases, we take a maximal false positive rate (FPR) of 3\% and 40\% (equivalent to minimum specificities of 97\% and 60\%), respectively, as our constraints. We then choose the optimal threshold that maximises the true positive rate (TPR) given these constraints. \cref{fig:ROC_usecases} shows the ROC curve obtained by our model, and reports the resulting TPRs, FPRs and decision thresholds $\hat{p}^t$. In the conservative case, the model could detect about two-thirds of diseased patients while only incorrectly classifying 3\% of healthy retinas as diseased. In the less conservative case, where we prioritise a high TPR over a low FPR, the model could detect 94\% of diseased retinas while incorrectly flagging up 40\% of healthy retinas as diseased. We thus conclude that our model would potentially be useful in clinical applications and performs very well overall, especially considering the more challenging test set compared to what had previously been considered.

\subsection{The proposed DL model performs excellently on a challenging external dataset}
\label{sec:main_externalval_results}
Following the example of recent work \citep{Antakibjophthalmol-2021-319030}, we assembled an external dataset of 75 UWF images. For a detailed description see \cref{sec:sup_exttestsetresults_detailed}. Though small, this external dataset is quite challenging in a number of ways: First, most images are taken with different types of UWF imaging devices (e.g. Optos\textcopyright \kern0.2em  Daytona or California) that produce qualitatively different images. Second, many of the images are cropped (and thus also have a different scale), projected, have a different aspect ratio or even contain watermarks. We did not correct for these factors through preprocessing. Third, these images are likely to not be of Japanese patients and thus present a different patient population. 

\begin{figure}[!th]
     \centering
    \includegraphics[width=0.65\textwidth]{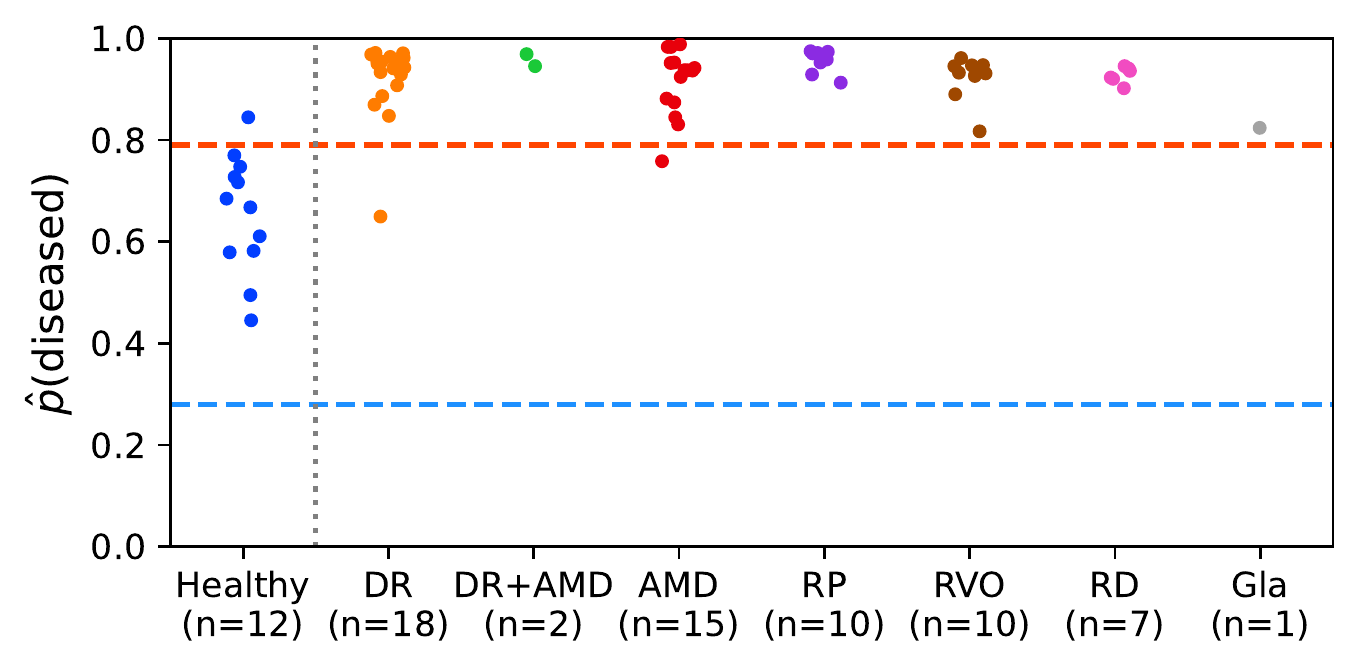}
    \caption{External validation results showing the predicted probability of being diseased $\hat{p}(\mathrm{diseased})$ stratified by ground-truth label. The red and blue horizontal lines plot indicate the conservative threshold $\hat{p}^t_\mathrm{conservative} =0.79$ and and less conserviative threshold $\hat{p}^t_\mathrm{less\kern 0.2em conservative} =0.28$, respectively.}
    \label{fig:external_validation_results_p_any}

\end{figure}

\cref{fig:external_validation_results_p_any} shows the predicted probability of being diseased stratified by ground-truth label. Despite these challenging conditions, our model achieved near perfect separation between healthy and diseased retinas (AUC=0.9841). Using the more conservative early screening threshold, our model would have had one False Positive, two False Negatives, and correctly classified all remaining 72 images. However, we also find that calibration of our model suffered under these conditions and using the less conservative threshold, we would have flagged up all images for investigation. This might be due all images looking potentially anomalous to our model as they are quite different from the images it was trained on. The model also identified the correct disease(s) in most cases, achieving very high label-wise AUCs for all included diseases (DR: 0.9627, RP: 1.0000, RVO: 0.9754, RD: 1.0000, Gla: 0.9595) apart from AMD where it achieved an AUC of 0.8051 which is good but noticeably worse than for the other labels. This might indicate that differences in imaging devices and data preprocessing have a larger impact on recognising AMD as such but it could also be due to differences in diagnostic thresholds.

We consider this to be a very encouraging result. Typically, we would expect worse results on external data especially if it differs from the training data as drastically as in this case. In practice, we would not recommend applying a model to images that are from different imaging devices or preprocessed in ways the model has not encountered during training. It is not unexpected that calibration would suffer due to this, but the excellent separation in terms of AUC is very promising. We would also like to note that we decided on the decision thresholds before the external dataset was collected, so it is remarkable that our model would have made only 3 mistakes out of 75 images using the conservative threshold.

\subsection{Regions that ML model pays attention to when diagnosing different diseases are consistent with domain knowledge}
\begin{figure}[!th]
     \centering
     \begin{subfigure}[b]{0.7\textwidth}
         \centering
         \includegraphics[width=\textwidth]{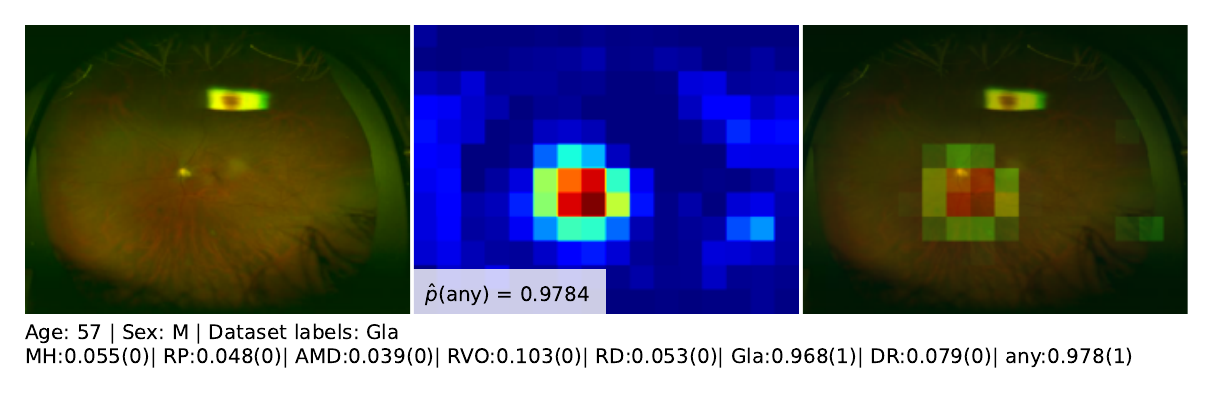}
         \caption{An example of the model successfully ignoring a reflection artefact and instead attending to the optic disc to correctly detect that the retina is diseased and that the disease in particular is Glaucoma ($\hat{p}(Gla)=0.968$).}
        \label{fig:imgwise_attention_heatmap_noisy_TruePositive_ignoringnoise}
     \end{subfigure}
     \begin{subfigure}[b]{0.7\textwidth}
         \centering
         \includegraphics[width=\textwidth]{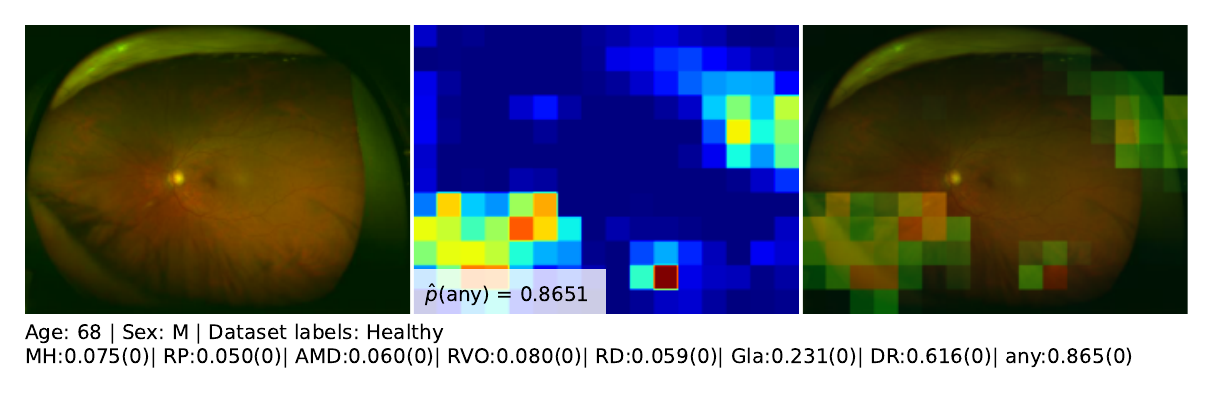}
         \caption{An example of the model focusing on artefacts, particularly eyelash artefacts, in the periphery and falsely predicting a high probability of being diseased. The predicted probability of DR is highest ($\hat{p}(DR)=0.616$) which could indicate that the model confuses the eyelash artefacts for haemorrhages or microaneurysms. There is also a small bright spot south-west of the optic disc which receives attention and might be interpreted as an exudate.}
        \label{fig:imgwise_attention_heatmap_noisy_FalsePositive}
     \end{subfigure}
        \caption{Examples of the model attention heatmaps generated by GradCAM. \protect\myunderline{Left}: Original image. \protect\myunderline{Middle}: Heatmap. \protect\myunderline{Right}: Heatmap imposed on the image. The text below the image indicates the patient's age, sex, and disease status according to the dataset labels. The second line of text indicates the model predictions for each label and the dataset labels in brackets.}
        \label{fig:imgwise_attention_heatmaps}
\end{figure}

We used explainable AI techniques to understand how the model makes its predictions. First, we use GradCAM \citep{selvaraju2017grad} to generate attention maps for a given input image and target disease. These highlight which regions of the image the model considers evidence for the respective disease. We find that the model generally pays attention to regions of pathology even in the presence of distractions like reflection artefacts (\cref{fig:imgwise_attention_heatmap_noisy_TruePositive_ignoringnoise}) but sometimes gets confused by these (\cref{fig:imgwise_attention_heatmap_noisy_FalsePositive}). Overall, these maps could aid clinicians in practice to understand the model’s predictions, to draw their attention to regions of interest on an individual patient basis or to identify if the model might be confused by an artefact. 

\begin{figure}[!th]
     \centering
    \includegraphics[width=0.65\textwidth]{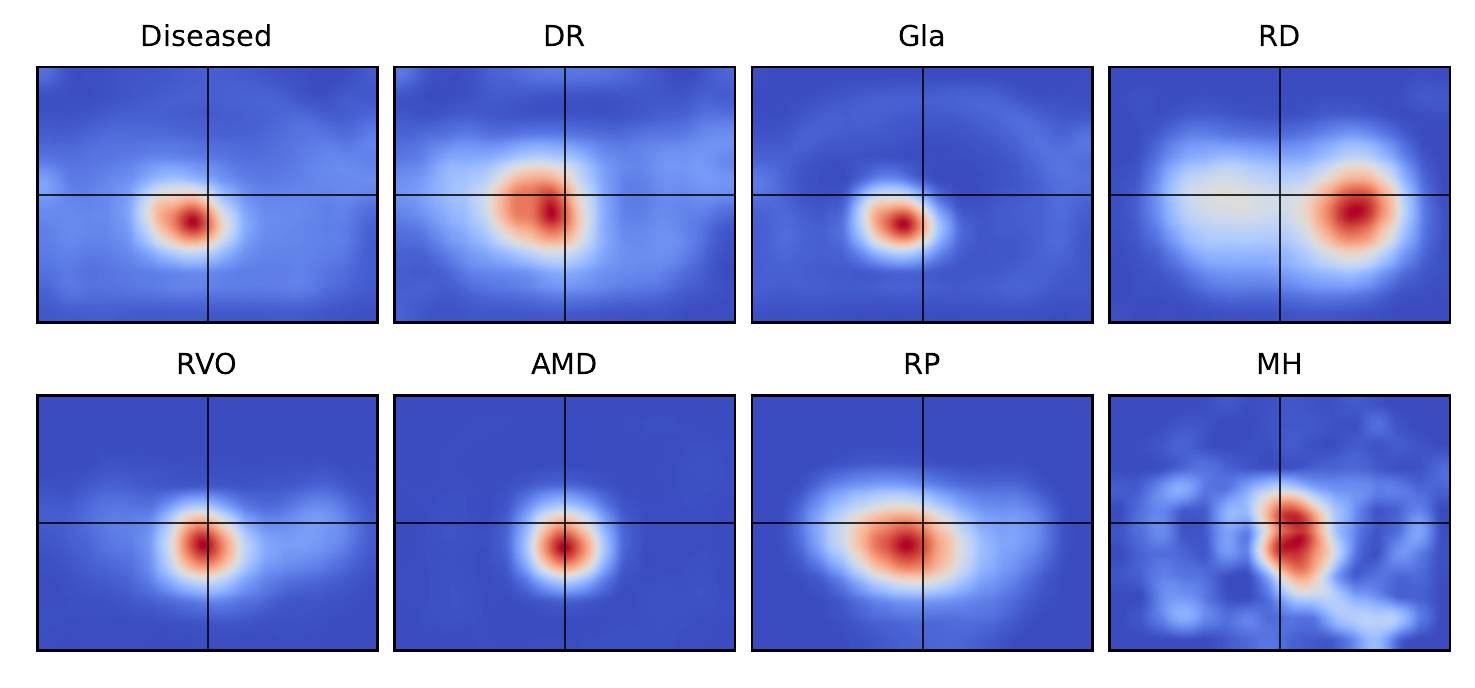}
    \caption{Label-wise global attention maps. Calculated on all images from the test set. Right eyes are flipped horizontally such that the temporal side is on the right for all images. Blue indicates regions of low attention, red regions of high attention. The centred black crosshairs are added to allow easier comparison of relative positions between subplots. }
    \label{fig:disease_average_heatmap}

\end{figure}
 
While individual image-wise attention maps are useful for understanding the model, only a small number of images can be examined in detail and reproduced at once, which constitutes a barrier to auditing DL models thoroughly and objectively. To examine the model more systematically and to understand which regions of the UWF images are useful to detect specific diseases, we generate label-wise global attention maps through predicted probability-weighted aggregation of image-wise attention maps \citep{global_explainability_our_arxiv_preprint}. \cref{fig:disease_average_heatmap} shows the resulting maps for the seven diseases and the general diseased label. These maps are consistent with domain knowledge. For example, the model focuses on the posterior pole for both Gla and AMD but for Gla this is concentrated on the optic disc side whereas for AMD this is concentrated on the macula. For RD, there is attention across the entire retina but concentrated on the temporal side where rhegmatogenous RD tends to occur most often \citep{wilkinson2017ryan}. 
For DR there also is attention across the retina but clearly concentrated around the optic disc, presumably due to cases of proliferate DR with neovascularization of the optic disc. Finally,  while the map for MH is concentrated on the macula it is also the noisiest map. This is due MH being the most difficult label for our model and due to the low number of images we average over as MH was the rarest disease in terms of images.
To summarise, these maps generally match domain knowledge. This indicates that the model detects pathology for specific diseases in regions where we would expect pathology to occur. This gives us confidence that the model works in a desirable fashion and does not leverage shortcut artefacts \citep{degrave2021ai}.

\subsection{Posterior pole region was identified as the most important region by the model and is sufficient for high performance}

\begin{figure}[!th]
     \centering
    \includegraphics[width=0.65\textwidth]{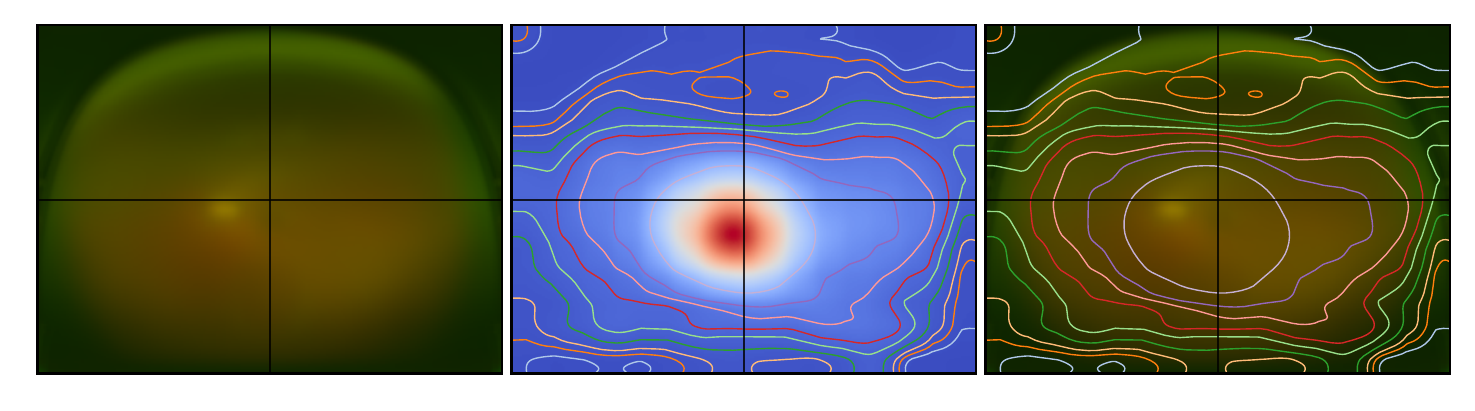}
    \caption{\protect\myunderline{Left}: Average of all test set images. Right eyes are flipped horizontally such that the temporal side is on the right for all images. \protect\myunderline{Middle}: Global attention map. Blue indicates regions of low attention, red regions of high attention. Contour lines indicate the most important regions in quantile steps of 10\%. \protect\myunderline{Right}: The average of all test set images with the same contour lines. The centred black crosshairs are added to allow easier comparison of relative positions between subplots. %
    }
    \label{fig:global_average_heatmap}

\end{figure}

We combined all label-wise global attention maps into a single global attention map (\cref{fig:global_average_heatmap}) \citep{global_explainability_our_arxiv_preprint}. This map ranks each position of the input images in terms of their overall importance to the model and identifies the posterior pole as the most important region. This agrees well with domain knowledge, but we would like to note that this was identified in a purely data-driven way. The global attention map also correctly identifies the corners which never show the retina as least important.

\begin{figure}[!th]
     \centering
    \includegraphics[width=0.99\textwidth]{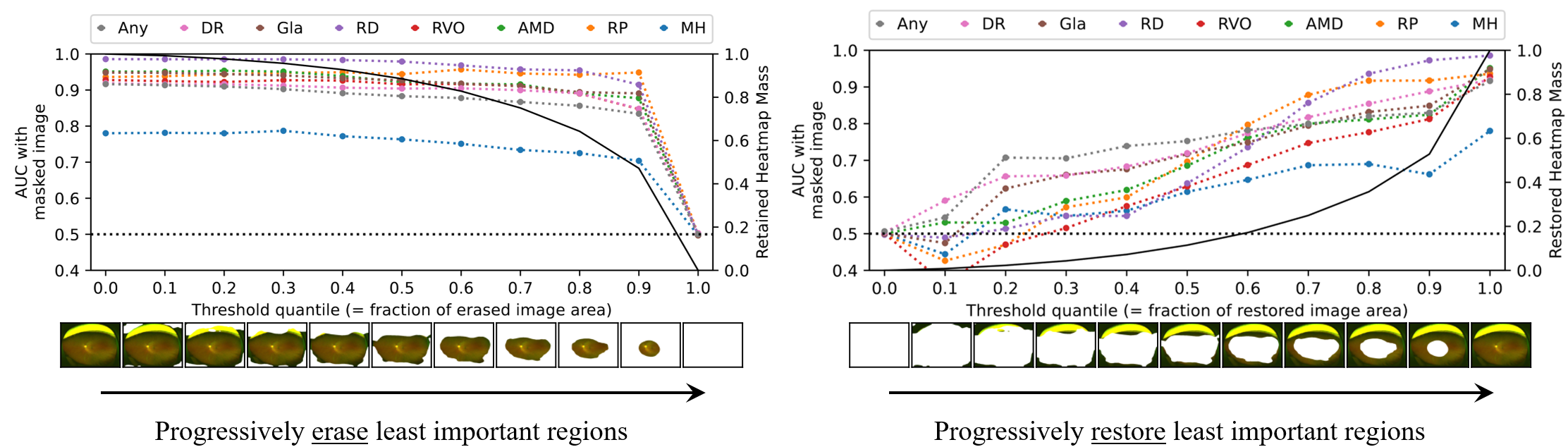}
    \caption{\protect\myunderline{Left}: Progressive erasure.
    \protect\myunderline{Right}: Progressive restoration.
    Coloured lines indicate the test set AUCs when erasing using the mask obtained with the respective threshold. The horizontal dashed black line indicates AUC=0.5 which is equivalent to random guessing. The solid black line indicates the mass of the global average heatmap at the respective threshold, using the secondary y-axis on the right. An example image with the mask at the given threshold applied is shown below the x-axes.
    }
    \label{fig:PEPPR_plots}

\end{figure}
We use Progressive Erasure Plus Progressive Restoration (PEPPR) \citep{global_explainability_our_arxiv_preprint} to validate that the global attention map is faithful to how the model makes its predictions and to investigate how the model’s performance degrades when removing either the most or the least important regions. \cref{fig:PEPPR_plots} shows the AUC for each label for the test set with different parts of the image removed.  When progressively erasing the least important image regions according to the global attention map, there is no significant drop in AUC even when 90\% of the image is removed. This indicates that this map does correctly reflect which regions are most important to the model. However, it is surprising that having only the most important 10\% of the UWF images is sufficient to obtain performance comparable to having the full images available. This suggests that all seven diseases have presentation in the posterior pole, which might be in part a reflection of the disease stages common in the TOP dataset. When progressively restoring the least important regions, starting with a blank image, we observe a near monotonic increase in AUCs with performance only peaking for all labels once the full image has been resorted. For diseases that primarily affect the optic disc and fovea (Gla, MH, RVO, and AMD), restoring the final most important 10\% containing the posterior pole leads to a large increase in AUC (> $\sim$ 0.1), whereas for diseases that cause more peripheral pathology (RD, RP and DR) there is a less substantial increase (< $\sim$  0.03). In summary, we find that the global attention map faithfully reflects how the model works. Surprisingly, the 10\% of the image containing the posterior pole are sufficient to achieve high performance.

\section{Discussion}
We find that DL is very effective for detecting disease in UWF images even when evaluating the model under more realistic conditions than what had previously been considered. This strong performance was also confirmed on a challenging external test set, including images taken with different cameras, different aspect ratios, or even watermarks. Thus, our DL model shows promise for applications such as early screening and clinical decision support. Using more realistic conditions (no artificial balancing of data, multiple diseases, comorbidities, no exclusion of difficult cases) for evaluation also resulted in lower AUCs than other published models \citep{matsuba2019accuracy, nagasato2019deep, nagasato2018deep, tabuchi2018discrimination, masumoto2018deep, nagasawa2018accuracy, nagasawa2019accuracy, ohsugi2017accuracy,masumoto2018retinal,masumoto2019accuracy}. To ensure that this is due to the test set being more challenging rather than our approach being inefficient, we compared our model against an ensemble of identical models that were trained as binary classifiers on individual diseases using artificially balanced data as is commonly done in the literature. We find that our approach clearly outperforms this approach both in terms of separation (AUC) and especially calibration (Brier score). For one previous study on the TOP dataset focused on RP \citep{masumoto2019accuracy}, the exact subset of the TOP dataset used after excluding images is available. On this subset, a simplified version of our model achieved perfect separation (AUC=1) with very high confidence in the correct labels (see \cref{sec:sup_RP_subset_previous_work}) which has also been noted in the literature \citep{Antakibjophthalmol-2021-319030}. Under more realistic and challenging conditions, on the other hand, our more complex model achieved an AUC of 0.9438 on the RP label which is still high but not perfect performance. Thus, we recommend using training and validation data that is as realistically as possible given the constraint of the available data, which has also been raised by clinicians \citep{tan2020deep}. While this might yield lower performance numbers, those are also more realistic and models trained on realistic data will fare better under realistic conditions.

We find that the model focuses on regions to diagnose specific diseases that are consistent with where we would expect pathology to occur. This concordance between domain knowledge and data-driven investigation implies that our model works in a desirable fashion and is unlikely to rely on “shortcut” artefacts which can be a problem in medical imaging \citep{degrave2021ai}. It also validates current domain knowledge, for instance we identify the posterior pole as the most important region purely from the data. Thus, this kind of approach might also be useful for knowledge discovery in the future. 

Surprisingly, the posterior pole region itself is sufficient to obtain high performance. This raises questions around whether and how all seven diseases we considered have presentation there. It also raises the question of how much benefit UWF imaging has over traditional CFP. It is possible that cases in the TOP dataset might skew towards being severe/progressed which can be detected in the posterior pole, whereas early signs might exclusively occur in the periphery. However, it has been noted that the TOP dataset contains both obvious and subtle cases of pathology at least for RD, RP, and RVO \citep{Antakibjophthalmol-2021-319030}. It is also possible that the DL model can spot subtle signs of early pathology in the posterior pole that have not been previously noted. DL has previously been shown to be able to extract information from fundus images that was not known to be present in those images, such as sex and cardiovascular disease risk \citep{poplin2018prediction,yamashita2020factors}. We also want to stress that AI-based diagnosis is far from the only use case of fundus imaging, and the additional retinal coverage of UWF images might help clinicians in making their own diagnosis, judging disease severity, and choosing appropriate interventions. 

Our work has several limitations, primarily due to constraints stemming from the available data. The TOP dataset is a very valuable resource, particularly due to its scale. However, there are signs of possible selection bias (see \cref{sec:sup_eda_TOP}), mislabelled examples, and some details are unclear, for instance how exactly labels were defined (e.g. thresholds for diseases like AMD and DR) and created (e.g. how much experience did the labellers have). Furthermore, though it does contain less obvious cases \citep{Antakibjophthalmol-2021-319030}, being a specialist clinic, the patients at Tsukazaki hospital might generally skew towards having severe cases. Presumably, most or all patients are Japanese and thus the population is relatively homogeneous in terms of genetics, culture-induced lifestyle factors, and access to healthcare. 

Future work could investigate in more detail how and where diseases present themselves in the posterior pole and whether the detection of less severe cases benefits more from having the periphery available. The latter could be achieved using the TOP dataset through manual curation of additional labels for disease severity by domain experts. Access to additional large labelled UWF datasets would allow for a more comprehensive external validation of our model.\footnote{We did contact authors of previous studies using private UWF datasets. Unfortunately, we were unable to access any such datasets thus far. We encourage the community to make datasets available whenever possible.} To evaluate the benefit of UWF over CFP in more detail, it would be particularly interesting to collect a dataset of CFPs and UWF images showing the same retinas at the same point in time to compare the two modalities directly. DL models designed for community screening should ideally be evaluated on datasets that closely resemble that use case, i.e. with  a large proportion of healthy images and where the diseased images are primarily early or mild cases. However, such UWF datasets are not currently available to the best of our knowledge.

It might be possible to improve on the performance of our proposed model, for instance by using extensive compute resources to manually improve the training procedure, do automatic hyperparameter tuning, or train multiple models for an ensemble. However, we think that the performance we obtained might be close to the saturation point for the TOP dataset. We examined the 20 most confident false positives of our model and found that at least 14 of them are likely pathological. For details see \cref{sec:sup_TOP20FPs_assessment}.
This also implies that our reported performance numbers underestimate the true model skill. Beyond outright label mistakes, with binary labels there is also conceptual ambiguity whether a borderline case is classified as pathological. Finally, improved technical performance on internal validation might not be particularly meaningful for clinical practice as the population a model encounters in practice will always be at least slightly different from the population the training data is from. This distribution shift is likely to wipe out marginal gains from extensive tuning.

The data-driven methods we used to identify informative regions could also be applied to CFP images, where to our knowledge previous work only defined regions of interest using domain knowledge in advance as opposed to taking a data-driven approach \citep{hemelings2021deep}. Furthermore, identifying informative regions in medical imaging could be combined with data augmentation methods such as MixUp \citep{zhang2017mixup} to allow for more data-efficient model training. 

We proposed a DL model and evaluated it under more challenging and realistic conditions than what had previously been considered. We find that the model is very effective at detecting diseased retinas and at identifying the correct disease(s). Thus, such models could be used in clinical practice. We further investigated which regions of the UWF image attends to and find that this matches domain knowledge. For instance, the posterior pole was identified as the most important region in a purely data-driven way. This indicates that the model works as intended. Surprisingly, using just the posterior pole region is sufficient to obtain high performance which should be investigated further in future research.

\section*{Acknowledgements}

We thank Dr. Hiroki Masumoto and all his colleagues at Tsukazaki hospital for releasing the TOP dataset. This is a great contribution to AI research in ophthalmology for which we are most grateful. 

We also thank the American Society of Retina Specialists for their Retina Image Bank, and RetinaRocks for their Image Library. We further thank all users that submitted images for research use to these online repositories or elsewhere.

Funding by the UKRI CDT in Biomedical AI is gratefully acknowledged.

\FloatBarrier
\section{Methods}
{\small
\singlespacing

\subsection{Detailed datasplit approach with patient-level stratification}
\label{sec:methods_datasplit_detailed}

We split the data at the patient- rather than the image-level. This is to avoid leaking information between the sets by having images of the same eyes or same patients across different sets. \cref{fig:example_mutipleimagesperpatient} shows an example of multiple images of the same eye in the TOP dataset. If we split at the image-level, images of this patient might end up in both the training and test set. A model that overfits (``memorises'') a patient's unique vasculature or the specific morphology of their pathology might then outperform another model that does not do this and would generalise better to unseen patients. With our approach, each patient occurs in exactly one of the three sets.

\begin{figure}[!tb]
\centering
\includegraphics[width=0.6\textwidth]{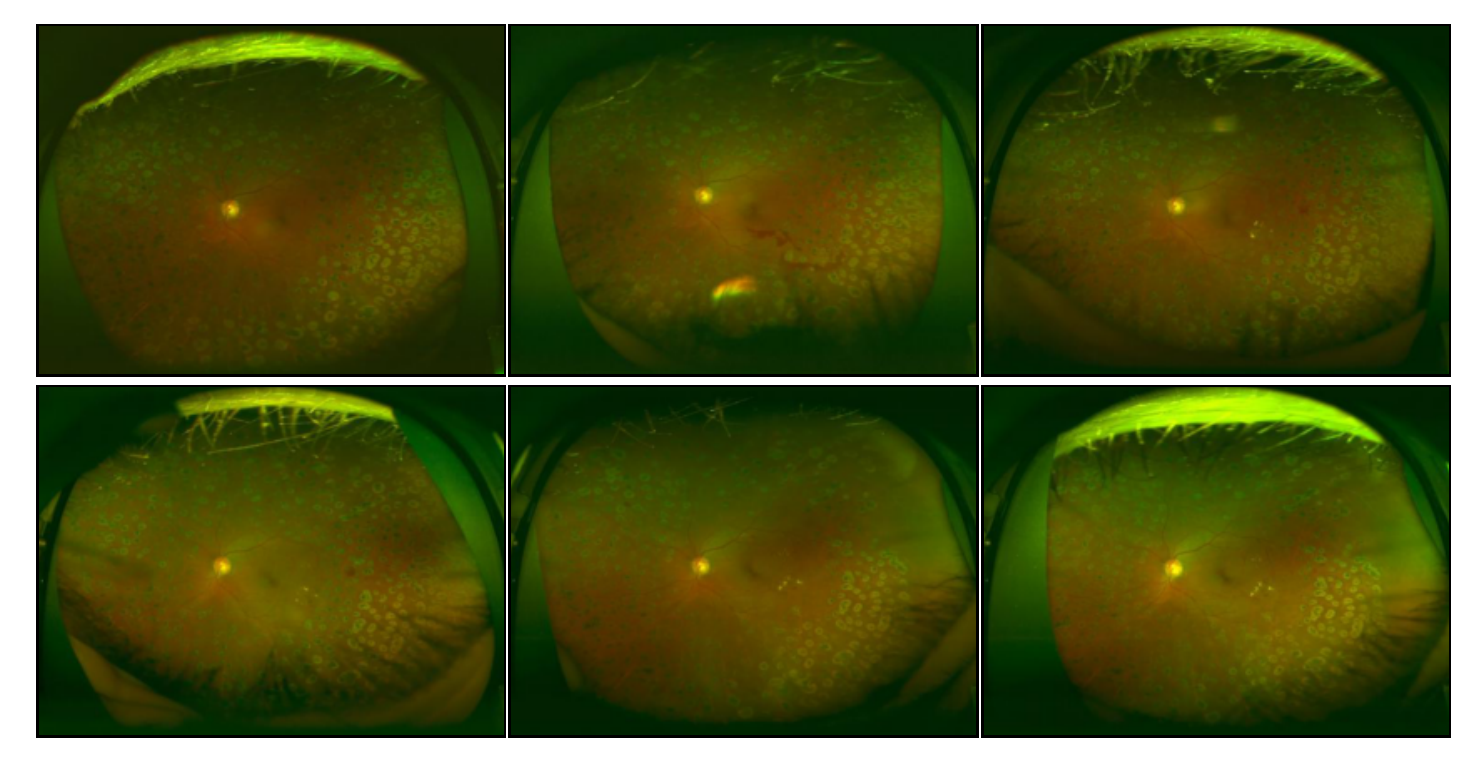}
\caption{Six of nine images showing the same eye of the same patient. All images show DR according to the labels. While there are some differences between the images in terms of artefacts and pathology, the general pattern of the pathology is consistent between images and could be memorised by a model.}
\label{fig:example_mutipleimagesperpatient}
\end{figure}

As some of the included diseases are relatively rare, we conduct a stratified split to ensure that the distribution of the diseases is as similar as possible across sets. However, as we have multiple, non-exclusive labels per image and multiple images per patient, stratification is more complex than if we had a single label per patient. Thus, we assign a stratification label to each patient according to the following three rules: First, if the patient has any disease across all their images, we take the disease that occurs most frequently as their stratification label. Second, if there is a tie (i.e. two or more diseases occur equally often in a given patient), we take the rarest of the diseases as their stratification label. Third, if and only if a patient has no disease in any of their images, we assign ``healthy'' as their stratification label. This gives us a single label with eight possible values, the seven diseases and ``healthy'', per patient and we can then stratify on that.

\subsection{Model and problem framing}
\label{sec:methods_model_problem_framing}

To develop a clinically useful model that can deal with comorbidities, we frame the problem as multi-label classification as opposed to binary\citep{matsuba2019accuracy, nagasato2019deep, nagasato2018deep, tabuchi2018discrimination, masumoto2018deep, nagasawa2018accuracy, nagasawa2019accuracy, ohsugi2017accuracy,masumoto2018retinal,masumoto2019accuracy} or multi-class classification\citep{Antakibjophthalmol-2021-319030} that have been used in prior work. The DL model outputs a probability for each label in both multi-class and multi-label classification. However, in the multi-class case, a softmax output activation is used which means that the model always outputs a total probability mass of 1 across all outputs, and thus an output for ``healthy control'' must be included. In the multi-label case, on the other hand, an element-wise sigmoid activation is applied to each output which allows the model to allocate a probability mass between 0 and 1 per label, and a total probability mass between 0 and $n_\mathrm{labels}$ across all labels. The sigmoid function $\sigma(x)=\frac{1}{1+e^{-x}}$ maps the raw model outputs $x\in\mathbb{R}$, called logits, to a predicted probability $\hat{p}\in (0,1)$.

Concretely, this allows the model to detect more than one disease in a given image. Suppose an image shows DR and AMD, then with an element-wise sigmoid activation the model can output high predicted probabilities $\hat{p}$ for both labels, whereas with a softmax activation it can only output moderate values of $\hat{p}$ for both classes or a high $\hat{p}$ for one class and a low $\hat{p}$ for the other. Likewise, suppose the model has only medium confidence that both of these diseases are present in the image. With an element-wise sigmoid, it can then output moderate values of $\hat{p}$ for both to express this. We use one label for each of the seven retinal diseases and also include a label for being ``diseased'' generally. This label allows the model to indicate that it is confident that a given image is affected by disease, whether it is confident in any disease in particular or not. Furthermore, this label is also the most clinically relevant label if we want to decide if a patient needs to be referred to an ophthalmologist for further examination. 

As our model, we use a convolutional neural network backbone which extracts a feature vector from an image and a prediction head which maps this feature vector to the eight-dimensional output to which we then apply element-wise sigmoid activations to obtain predicted probabilities for each label. We use a ResNet34 \citep{he2016deep} as convolutional backbone which consists of 33 convolutional layers with residual connections followed by an average pooling layer and a linear output layer which we replace with our own prediction head. After the last convolutional layer, the ResNet34 outputs a three-dimensional feature map, two spatial image dimensions and 512 channels, which the average pool converts to a flat 512-dimensional feature vector by averaging across the spatial dimensions. We use the ResNet34 in its original configuration with batch normalisation \citep{ioffe2015batch} and Rectified Linear Units (ReLU) as activation function where $ReLU(x)=max(0,x)$. We chose a ResNet as it is a high-performance, efficient and well-established architecture that has recently been shown to be very competitive when using modern training techniques \citep{wightman2021resnet, bello2021revisiting}, and ResNet34 in particular as it was the largest ResNet variant that fit into GPU memory at a reasonable batch size. We initially experimented with ResNet18 and found that moving to ResNet34 offered a very minor performance gain on the validation set. Thus, we expect that using larger models would not yield a significant performance improvement given that the dataset is small by deep learning standards.

Our prediction head is a feed-forward neural network with one hidden layer containing 128 hidden units and eight outputs, one for each label. As activation function for the hidden units in the prediction head, we use Parametric Rectified Linear Units (PReLU) \citep{he2015delving}  $PReLU(x)=max(0,x)+\alpha  min(0,x)$, which extends ReLU by not zeroing out negative inputs completely which improves convergence and where the negative slope parameter $\alpha$ is itself a learnable parameter. We learn a separate $\alpha$ for each unit of the hidden layer. Using a prediction head with a hidden layer rather than linear head allows our model to more easily learn appropriate correlations between labels. Our entire model thus is a 35-layer deep neural network with 21,348,360 trainable parameters.

\subsection{Model training}
\label{sec:methods_training}

We initialise the weights of the convolutional ResNet34 layers with weights trained on ImageNet \citep{deng2009imagenet}, a large dataset of natural images. This practice is commonly called transfer learning. However, since UWF retinal images are quite different from natural images, the pretrained weights can be thought of as a refinement of random initialisation where the weights between layers are tuned to each other and potentially useful filters in lower levels have already been learned. To adapt the weights of the input layer from 3-channel natural to 2-channel UWF images, we take only the weights for the first two channels and weigh them by 3/2 to preserve the input layer's activation mass.\footnote{This is the default behaviour of the PyTorch Image Models library (timm) and described in more detail here: \url{https://fastai.github.io/timmdocs/models}.} The linear layers of the prediction head are initialised using uniform Kaiming initialisation \citep{he2015delving} with $a=\sqrt{5}$ and  PReLU $\alpha$s are initialised to 0.25, both of which are the default initialisations used by PyTorch. %

Our neural network model is a function $f$ parameterised by model parameters $\theta$ (weights, biases, batchnorm affine parameters and PReLU $\alpha$s) that maps the image space $\mathbb{R}^\mathrm{HxWxC}$ with height H, width W and pseudocolor-channels C=2 to the eight-dimensional label space $f_\theta : \mathbb{R}^\mathrm{HxWxC} \to  \mathbb{R}^8$.
For an individual image $X_i$, we can then interpret the $l$-th element of the model output $f_\theta(X_i)_l=\hat{p}_{i,l} \in (0,1)$ as a confidence score that $X_i$ shows the disease indicated by the $l$-th label. We fit the parameters $\theta$ to minimise the total label-wise logistic loss (also referred to as binary crossentropy) across for all training images $i \in n_\mathrm{images}^\mathrm{train}$. Thus, our objective is 
\[ \min_\theta \sum_{i=1}^{n_\mathrm{images}^\mathrm{train}} \sum_{l=1}^{n_\mathrm{labels}} -(y_{i,l} \log (f_\theta(X_i)_l) + (1 - y_{i,l}) \log (1 - f_\theta(X_i)_l))\]
where $y_{i,l} \in \{0,1\}$ is the true value for the $l$-th binary label of the $i$-th image. 

We optimise this objective using mini-batch Stochastic Gradient Descent (SGD) with momentum $\gamma=0.9$. The learning rate $\eta$ is set dynamically using a cosine annealing schedule with warm restarts \citep{loshchilov2016sgdr} which starts with a high learning rate that is then decayed over time and reset once the minimum of the cosine function has been reached. The learning rate of a given epoch $\eta_t$ is given by $\eta_t = \frac{1}{2}\left(1 +\cos\left(\frac{T_{cur}}{T_{i}}\pi\right)\right)\eta_{max}$ where $T_i$ is the number of epochs between warm restarts and $T_{curr}$ is the number of epochs since the last restart. Once $T_{curr}=T_i$, $T_{curr}$ is reset to 0, which implies  $\eta_t=\eta_{max}$. We use maximum learning rate $\eta_{max}=0.1$ and $T_i$ = 10. The learning rate is updated after each mini-batch, thus $T_{curr}$ can take on fractional values. We train for 30 epochs with mini-batches of size 32, which was the largest size that fit our available GPU memory, shuffling the data before each epoch. Initially, we trained for 100 epochs using a constant learning rate, but after switching to a cosine schedule our model consistently converged within 30 epochs.

To make our model converge more smoothly and to better performance, we use a technique called Model Exponential Moving Average (EMA).\footnote{Averaging over the trajectory in stochastic optimisation has its roots in Polyak-Ruppert averaging. Other forms of averaging have been explicitly proposed in the context of modern deep learning \citep{izmailov2018averaging}. However, EMA specifically appears to be a trick that is used in the literature and supported by major libraries (e.g. tensorflow \nolinkurl{https://www.tensorflow.org/api_docs/python/tf/train/ExponentialMovingAverage} and timm \nolinkurl{https://fastai.github.io/timmdocs/training_modelEMA}) yet has not been introduced formally by a canonical work.} At the start of training, we create a copy of the model parameters $\theta$ called $\theta_\mathrm{EMA}$. After each mini-batch and update to $\theta$, we update the current EMA model parameters $\theta_\mathrm{EMA}^t$ to a weighted average of the previous EMA model parameters and current model parameters  $\theta_\mathrm{EMA}^t=\alpha \theta_\mathrm{EMA}^{t-1} + (1-\alpha) \theta$ where $\alpha$ is a decay hyperparameter, which we set to $0.999$. At the end of training, we use $\theta_\mathrm{EMA}$ as the final model parameters. 
When not using EMA, the label-wise validation AUCs have high variance across epochs as AUC is a ranking metric, but with EMA they converge smoothly and to higher values.

Overfitting is a key problem in machine learning, especially when fitting large deep learning models with millions of parameters. This is when our model successfully minimises our loss function for the training data but performs poorly on unseen data. We want the model to learn to identify generalisable patterns relevant to the target labels, for instance to look for drusen to decide whether the image shows AMD. However, the model can also minimise the loss function for the training data by ``memorising'' hyper-specific patterns of individual training examples (e.g. specific eyelash artefacts or non-pathological, distinctive vasculature of a patient) and their corresponding labels. Such an overfit model works well on the training data but would be ineffective for diagnosing unseen images. Overfitting can be addressed through regularisation which makes it harder to focus on such hyper-specific patterns and encourages the model to learn patterns that generalise instead. We find that in this application, even a smaller model with a ResNet18 as backbone can fit the training data perfectly (label-wise AUCs equal or almost equal to 1) after a few epochs when not applying any regularisation.

A key regularisation technique for computer vision is data augmentation, where images are randomly perturbed during training to yield images that retain most of their characteristic, generalisable patterns but where non-generalisable, hyper-specific patterns are harder to recognise. In our application, data augmentation is especially important given that our dataset contains a few thousand images. This is large by biomedical standards yet small by modern deep learning standards where datasets can contain hundreds of millions of images.\footnote{E.g. \url{https://paperswithcode.com/dataset/jft-300m}.}

The augmentations that we use can be split into three types: flip, domain, and general augmentations. As many patterns of pathology, such as drusen, are rotation invariant, we randomly flip the images, which are all read as left eyes, horizontally with probability $p^\mathrm{hflip}=0.3$ and vertically with $p^\mathrm{vflip}=0.1$ which increases diversity while still leaving most images in the orientation we will use when predicting on unseen images. Our domain augmentations emulate some common variations in UWF image quality. We randomly scale all values in either channel with a separate float drawn uniformly from the interval $[0.75, 1.25]$ to vary both brightness and contrast. We further apply a Gaussian blur kernel with kernel size 7x7 and $\sigma$ drawn uniformly from $[0.1,1]$ to emulate out of focus images. Finally, we add Gaussian noise individually drawn from $\mathcal{N}(0, 0.1)$ to each pixel of either channel to emulate noise introduced by media opacity. All domain augmentations are applied jointly with $p=0.9$. We also use some general augmentations that are commonly used in machine learning. We use RandomErasing \citep{zhong2020random} which with $p=0.5$ replaces a rectangle with a fraction of the overall image area drawn uniformly from $[0.05, 0.3]$ and aspect ratio drawn from $[0.3, 3.3]$ with noise drawn from a standard normal distribution. We further apply a random affine transformation which rotates the image by an angle drawn uniformly from $[-15,15]$ degrees, scales it with a factor drawn uniformly from $[0.8, 1.2]$ and shears it by an angle drawn uniformly from $[-10,10]$ degrees. These general augmentations are also applied jointly with $p=0.9$ where RandomErasing is then applied with $p=0.5$.

Further, we use Mixup \citep{zhang2017mixup} during training. Mixup takes two images $X_i$ and their respective target binary label vectors $y_i$ and blends them together with linear interpolation as a new datapoint $\{X^\mathrm{Mixup}=\lambda X_1 + (1-\lambda) X_2, y^\mathrm{Mixup}=\lambda y_1 + (1-\lambda) y_2\}$. We draw the blending parameter $\lambda$ from $\lambda \sim \beta (0.6,0.6)$, which means that most values of $\lambda$ are close to 0 or 1. The result is an image where one of the images can be seen clearly with the other one being faint, and the label vector being weighted accordingly. For instance, suppose we combine a healthy control and an image showing AMD with $\lambda=0.9$, then the resulting image will primarily look like the healthy control with the AMD image faintly showing through, including the corresponding pathology like drusen. The target labels will be $0.9\times0+0.1\times1=0.1$ for AMD and ``any retinal disease''. Mixup reduces memorisation and improves model performance in the presence of mislabeled examples and its resistance against adversarial examples \citep{zhang2017mixup}. During training, we pair up each input of a mini-batch with another and generate two blends, one with either input being in the first position.

Finally, we use three more regularisation techniques to improve model performance. First, we apply Dropout \citep{srivastava2014dropout} in our prediction head. Dropout regularises the model by randomly zeroing out hidden units during training and reweighing the remaining activations to preserve mass. This prevents undesirable co-adaptation between hidden units and improves generalisation. When using Dropout, we can interpret the final model during inference as an ensemble of many submodels. We use Dropout after the pooling layer with a low probability of $p=0.1$  to encourage the model to not rely on only a few of the 512 extracted features without forcing the convolutional model to output redundant features, and after the hidden layer with $p=0.5$ which yields the greatest diversity of submodels. Second, we use a small weight penalty of $\beta=0.5\times10^{-6}$ which penalises the $L2$-norm of the model parameters $\theta$ thus changing the objective to $\min_\theta L_\mathrm{original} + \beta \Vert \theta \Vert_2$ where $L_\mathrm{original}$ is the original logistic loss to encourage lower magnitudes of each parameter which reduces overfitting. Third, we use Label Smoothing (LS) \citep{szegedy2016rethinking}. LS replaces binary target labels with soft targets. This prevents the model from outputting overly confident predictions and instead encourages it to map all samples where it is highly confident to the same value which improves model calibration \citep{muller2019does}.  We use asymmetric LS, replacing 1s with 0.99 and 0s with 0.05. This allows the model to be confident in a diagnosis while discouraging it from entirely discounting the possibility of rare disease occurring. LS can also be interpreted as encoding a belief about the possibility of images being mislabeled;we consider it more likely that a diseased image is falsely labelled as healthy control, with all labels being 0, than vice versa. 

\subsection{Image preprocessing}
\label{sec:methods_image_preprocessing}
The images are shared in JPEG format at a resolution of 768x1024 pixels. This is a lower resolution than the scanner acquires but still provides a detailed picture of the retina. The scanner only acquires two channels but the JPEGs also contain a third channel with low, predominantly zero values which we assessed to have no important information. JPEG compression adds a third channel and introduces such cross-channel artefacts, and so this channel was discarded from consideration. In practice, it would be ideal to input the scans into the DL model in a lossless format, in which case no third channel would be present. We thus remove this channel for the present work and input a 2-channel image into the model.

We downscale images to 384x512, reducing the number of pixels by a factor of 4. As the input resolution affects the size of the convolutional feature maps, this drastically lowers the GPU memory usage. We flip all right eyes horizontally, so that all images are approximately aligned. %

\subsection{Benchmark models}
\label{sec:methods_baselines}
For the non-image baseline using age and sex (the only available patient covariates), we considered three different classification algorithms: Logistic Regression, RandomForestClassifier (RFC) \citep{breiman2001random}, and kNearestNeighbours (KNN).\footnote{Logistic Regression and KNN are standard methods and described in textbooks such as \citep{friedman2001elements}.} As we tuned the hyperparameters of our main model, we also tune the hyperparameters of this baseline for a fair comparison. See \cref{sec:sup_baselinehyperparams} for details.

For the Ensemble of Experts, we used the same architecture and training schedule as for our main model. However, as subsampling for rarer diseases leads to small datasets, we evaluated both the EMA and non-EMA model parameters on the validation set and picked the better performing parameters for the test set evaluation. To obtain predictions for the general ``diseased'' label from the expert binary models for each disease, we take a summary statistic of the disease-wise predictions of all expert models. We tried the minimum, mean, median and maximum and found that the maximum performed best on the validation set. For reference, using the maximum of the disease-wise predictions instead of the prediction for the ``diseased'' label for our proposed DL model leads to a minor drop in performance from AUC=0.9206 to 0.9103.

\subsection{Progressive Erasure Plus Progressive Restoration (PEPPR)}
\label{sec:methods_peppr}

For PEPPR \citep{global_explainability_our_arxiv_preprint}, we replaced erased pixels with noise drawn from a standard normal distribution so that this mirrors the RandomErasing data augmentation. Thus, our model encountered similarly erased regions during training. Furthermore, we note that we did not retrain our DL model at each step. The reported results are obtained with the weights of our final proposed DL model. This is because we want to audit what this specific model has learned and to validate that it did not make use of any shortcut artefacts \citep{degrave2021ai}. Even though the model was trained on full images, it is still able to achieve very strong performance with only the central 10\% of the image showing the posterior pole. 
To avoid data leakage, we use the global attention map obtained on the validation set, so that no test set information is used to select the regions for erasure which could, in principle, make the global attention map appear more accurate that it is. However, in practice, the global attention map for the validation and test sets look virtually identical. 

\subsection{Implementation}
The code for this project was implemented in Python 3.8.8 and is available at \url{https://github.com/justinengelmann/UWF_multiple_disease_detection}.
We used PyTorch \citep{NEURIPS2019_9015} version 1.8.1 and the PyTorch Image Models (timm) \citep{rw2019timm} library version 0.4.9 for our deep learning models. Particularly, timm was used for pretrained model weights and for Model Exponential Moving Average and Mixup. For Mixup, we make a minor modification to support the multi-label case rather than the multi-class case, which we include in the code files. For general scientific computing we used NumPy \citep{harris2020array} version 1.20.1 and for non-deep learning classification algorithms, metrics and other utility code we used scikit-learn \citep{scikit-learn} version 0.24.2. Plots were generated with the Matplotlib \citep{Hunter:2007} and seaborn \citep{Waskom2021} libraries.

} %

{
\singlespacing
\small
\bibliography{thesis}
}
\FloatBarrier
\newpage
\section{Supplementary Materials}

\FloatBarrier
\subsection{Exploratory data analysis of the TOP dataset}
\label{sec:sup_eda_TOP}

The TOP dataset contains a label for diabetes mellitus (DM) as determined by a blood test with no further details given. We find that this label and DR co-occur often (Dice coefficient of 0.09). However, in general not all patients with DM will have DR, so the high co-occurrence in the TOP dataset might indicate that a majority of DM patients were examined because of suspected DR or that DM blood tests were done primarily for patients showing signs of DR. Curiously, 39 images showed DR but the patient did not have DM according to the labels, which supports the idea that
the DM label refers only to blood tests done at Tsukazaki hospital and thus there might be patients with DM who are recorded as DM negative in the dataset. 

The equal sex balance which is surprising given that patients in the TOP dataset tend to be of advanced age and that females tend to live longer. One possible explanation could be a conscious decision by the researchers at Tsukazaki hospital to include an equal proportion of males and females. Thus, it could be a sign of possible selection bias.

\begin{figure}[!ht]
     \centering
     \begin{subfigure}[b]{0.49\textwidth}
         \centering
         \includegraphics[width=\textwidth]{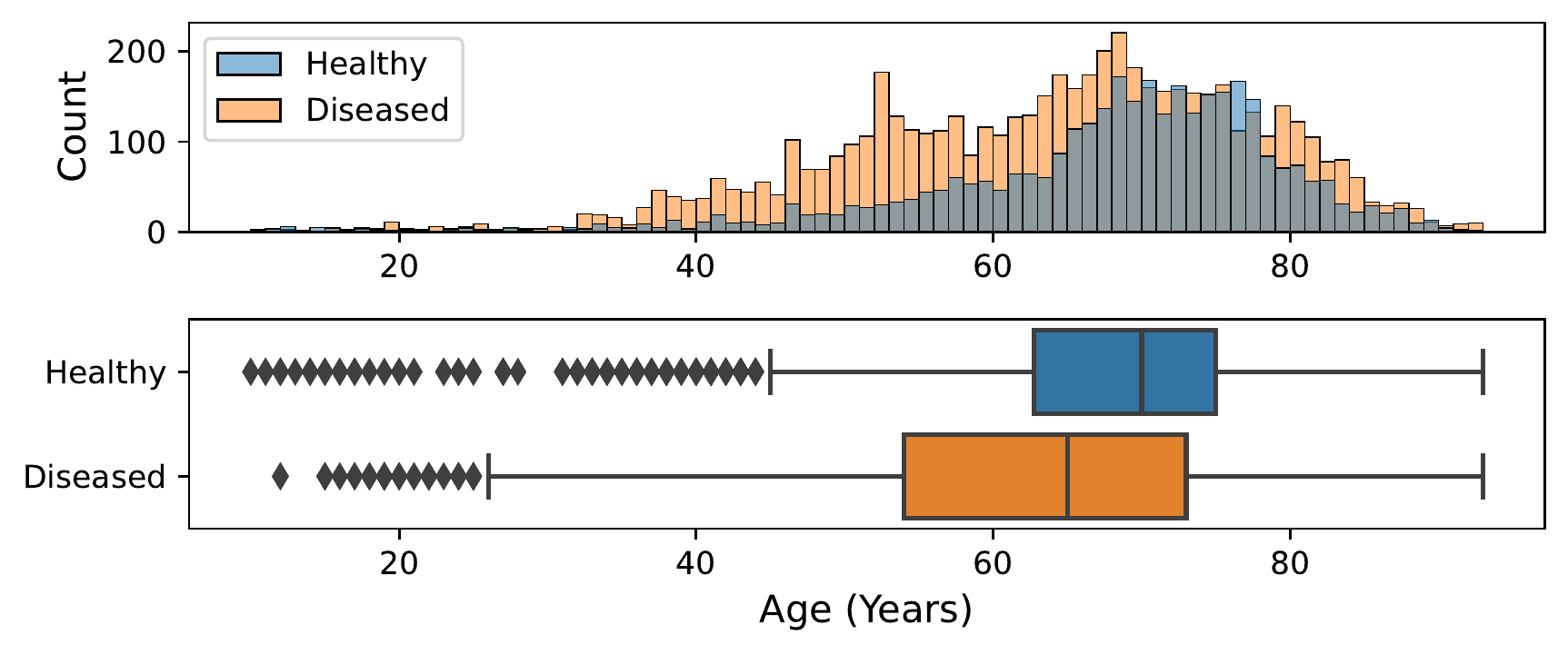}
         \caption{Image-level (n=9121)}
     \end{subfigure}
     \hfill
     \begin{subfigure}[b]{0.49\textwidth}
         \centering
         \includegraphics[width=\textwidth]{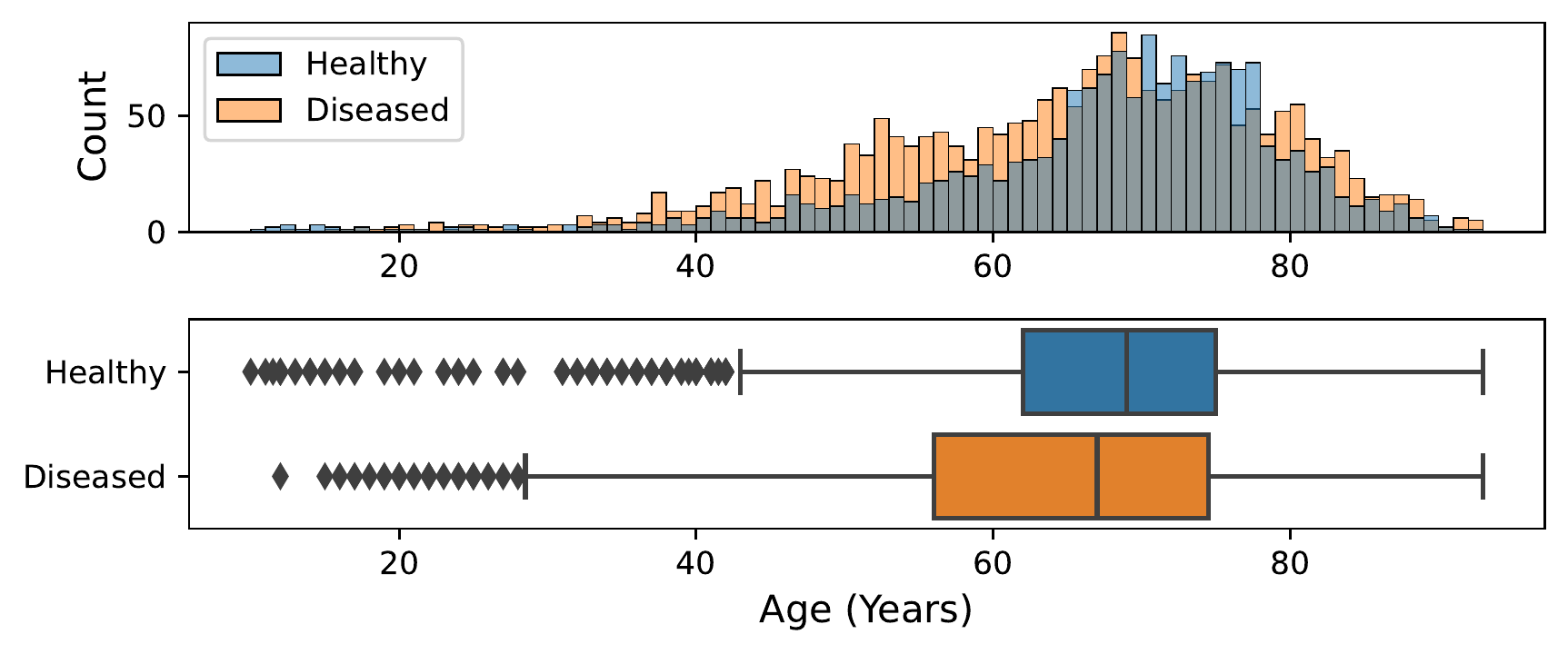}
         \caption{Patient-level (n=3763)}
     \end{subfigure}
        \caption{Distribution of patient age stratified by disease status for the train set. Bin width is set to 1, the granularity of age in the dataset. To define the disease status at the patient-level, we classify a patient as diseased if any of their images showed a retinal disease or if they have DM.}
        \label{fig:age_distribution}
\end{figure}

More detailed analysis was done on the train set only after the data was split. This is to ensure that we do not leak information from the test set into our modelling process.  \cref{fig:age_distribution} shows the distribution of age stratified by disease status. There is some coverage of almost all age groups, from young children to centenaries. Most patients are between 50 and 80 which is likely reflective of the fact that most retinal diseases occur in older patients and of Japan's demographic makeup. Surprisingly, we find that healthy patients are generally older than diseased patients. This difference is more apparent at the image-level but persists at the patient-level, even when using the sweeping definition of classifying a patient as diseased if they had any disease associated with any of their images. We think that this is another sign of possible selection bias. Patients get examined at a specialist clinic for a reason. With younger patients, this would be concrete reasons to suspect retinal diseases like self-reported deteriorating vision, whereas older patients might get examined routinely due to their age, even if they have no symptoms.

Another possible sign of selection bias is that the labels for DM and DR have low co-occurrence with the other retinal diseases in the TOP dataset. We would generally expect that patients with diabetes have an increased risk of many other conditions, including retinal diseases other than DR. However, looking at the TOP dataset, patients with DR/DM appear to have a much lower risk of having other retinal diseases than patients without DR/DM. This could again be due to patients being referred to Tsukazaki hospital for a reason, which could be suspected DR or some other reason. Patients that do not have DR/DM are more likely to have been referred for some other reason and thus have other retinal diseases more often.

\FloatBarrier
\subsection{Overview of work by researchers from Tsukazaki hospital}

\begin{table}[!ht]%
\caption{An overview of prior work on the TOP dataset conducted by researchers from Tsukazaki hospital. ``Custom 1'' is an architecture consisting of 3 convolutional layers, followed by a maxpool and a linear prediction head. ``Custom 2'' consists of 3 convolutional layers each followed by a maxpool, followed by a prediction head containing one hidden layer. ``Ensemble'' refers to an ensemble of multiple convolutional neural networks. Where multiple tasks where considered (e.g. distinguishing a subtype from healthy controls), we selected the most relevant performance metric. One work did not report AUC, so sensitivity and specificity are reported instead.}
\label{tab:tsukazaki_litreview}
\centering
\begin{adjustbox}{max width=\textwidth, max totalheight=1\textheight-2\baselineskip}
{\small
\begin{tabular}{@{}llllllll@{}}
\toprule
Reference & Disease & Diseased & Controls & Reported Performance                                                           & Model            & Validation & GradCAM? \\ \midrule
\cite{matsuba2019accuracy} & AMD & 137 & 227  & AUC=0.9976 & Custom 1 & 70-30 split & Yes \\
\cite{nagasato2019deep} & RVO & 237 & 229  & AUC=0.976  & VGG16    & k-fold CV   & Yes \\
\cite{nagasato2018deep} & RVO & 125 & 238  & AUC=0.989  & VGG16    & k-fold CV   & Yes \\
\cite{tabuchi2018discrimination} & Gla & 950 & 1677 & AUC=0.987  & VGG16    & k-fold CV   & No  \\
\cite{masumoto2018deep} & Gla & 982 & 417  & AUC=0.872  & Custom 1 & 80-20 split & Yes \\
\cite{nagasawa2018accuracy} & MH  & 195 & 715  & AUC=0.9993 & Custom 1 & 80-20 split & Yes \\
\cite{nagasawa2019accuracy} & DR  & 132 & 246  & AUC=0.969  & VGG16    & k-fold CV   & Yes \\
\cite{ohsugi2017accuracy} & RD  & 411 & 420  & AUC=0.988  & Custom 2 & 75-25 split & No  \\
\cite{masumoto2018retinal}          & RD      & 600      & 818      & \begin{tabular}[c]{@{}l@{}}Sensitivity=0.973,\\ Specificity=0.915\end{tabular} & Ensemble & k-fold CV  & No       \\
\cite{masumoto2019accuracy} & RP  & 150 & 223  & AUC=0.998  & VGG16    & k-fold CV   & Yes \\ \bottomrule
\end{tabular}

}
\end{adjustbox}
\end{table}%

\FloatBarrier
\subsection{Details of the external test set}

\cref{tab:sup_exttestsetresults_detailed} lists the sources, filenames, labels and predicted probabilities for all images of the external test set. Using these data sources to assemble an external test set was inspired by recent work by \cite{Antakibjophthalmol-2021-319030}. ``ORP'' refers to Optos'{\textcopyright}  Recognizing Pathology which contains UWF images with clinical labels designed as ``a searchable reference resource to support clinical decision making''. ``ASRS'' refers to the American Society of Retina Specialists'  Retina Image Bank® and ``RetinaRocks'' for the RetinaRocks Image Library. 

For images where no clear label is provided at the source we take the labels from \cite{Antakibjophthalmol-2021-319030} who are trained ophthalmologists. We thank everyone who made these images available for research use, for which we are most grateful. We do not reproduce any images here, but we note that the American Society of Retina Specialists requests the following acknowledgement for each image: ``This image was originally published in the Retina Image Bank® website. © the American Society of Retina Specialists.'' The exact submitter for each image from that database can be found at the respective URL. We do not print the full links here due to space constraints, but they are available as hyperlinks in the ``Source/URL'' column. RetinaRocks images are provided in a Google Drive and thus cannot be linked directly. The filename allows to identify the exact image in this case. 

We were unable to find external UWF images that were described as showing a Macular Hole. For Glaucoma, we were only able to find one additional image. In the future, we hope to be able to test our model on larger external datasets.

For stereo images from ORP, we follow \cite{Antakibjophthalmol-2021-319030} in taking the left of the two panels to avoid cherry-picking. Some images from RetinaRocks had both eyes side by side in a single image. For those, we simply manually split them into two separate files. We applied our standard data pipeline to all external images: We resized them to the same resolution we used for the TOP dataset, removed the third channel present in JPG images, and flipped right eyes horizontally. No further processing was done, e.g. to correct for the different scale of cropped images, different aspect ratios, or watermarks. Together with the fact that many of these images were taken with different UWF scanners than what was used at Tsukazaki hospital, this makes the external dataset a very challenging stress test for our model. Furthermore, we want to note that any particularities in the data collection that might be present in the TOP dataset will also be absent from these images taken from a wide variety of sources. Thus, good performance on this dataset would indicate good generalisation of a model.

\label{sec:sup_exttestsetresults_detailed}
\begin{table}[!tp]%
\caption{Details of the external test set.}
\label{tab:sup_exttestsetresults_detailed}
\centering
\begin{adjustbox}{max width=\textwidth, max totalheight=1\textheight-2\baselineskip}
{\small
\begin{tabular}{rlllr}
\toprule
 Image &           Label &                                                                                                                                                         Source/URL &                                                                              Filename &  $\hat{p}(diseased)$ \\
\midrule
     1 &             AMD &                         \href{https://recognizingpathology.optos.com/wp-content/uploads/2016/02/Color-Atrophic-AMD-with-Geographic-Atrophy-OS-California.jpg}{ORP} &                          Color-Atrophic-AMD-with-Geographic-Atrophy-OD-California.jpg &     0.982783 \\
     2 &             AMD &                                                    \href{https://recognizingpathology.optos.com/wp-content/uploads/2016/02/Color-Dry-AMD-OD-California-1.jpg}{ORP} &                                                     Color-Dry-AMD-OD-California-1.jpg &     0.937082 \\
     3 &             AMD &                                                    \href{https://recognizingpathology.optos.com/wp-content/uploads/2016/02/Color-Dry-AMD-OS-California-1.jpg}{ORP} &                                                     Color-Dry-AMD-OS-California-1.jpg &     0.941377 \\
     4 &             AMD &                                                                   \href{https://recognizingpathology.optos.com/wp-content/uploads/2016/02/California-AMD.jpg}{ORP} &                                                                    California-AMD.jpg &     0.873758 \\
     5 &             AMD &                                                               \href{https://recognizingpathology.optos.com/wp-content/uploads/2016/02/California-AMD-Dry.jpg}{ORP} &                                                                California-AMD-Dry.jpg &     0.923850 \\
     6 &             AMD &                                                                 \href{https://recognizingpathology.optos.com/wp-content/uploads/2016/02/California-AMD-3.jpg}{ORP} &                                                                  California-AMD-3.jpg &     0.951265 \\
     7 &             AMD &                                                                 \href{https://recognizingpathology.optos.com/wp-content/uploads/2016/02/California-AMD-2.jpg}{ORP} &                                                                  California-AMD-2.jpg &     0.982928 \\
     8 &             AMD &                                                   \href{https://recognizingpathology.optos.com/wp-content/uploads/2016/02/Daytona-Projected-AMD-2-of-2-1.jpg}{ORP} &                                                    Daytona-Projected-AMD-2-of-2-1.jpg &     0.881232 \\
     9 &             AMD &                                                     \href{https://recognizingpathology.optos.com/wp-content/uploads/2016/02/Daytona-Projected-AMD-1-of-2.jpg}{ORP} &                                                      Daytona-Projected-AMD-1-of-2.jpg &     0.952176 \\
    10 &             AMD &                                                      \href{https://recognizingpathology.optos.com/wp-content/uploads/2016/02/Daytona-Projected-Wet-AMD-1.jpg}{ORP} &                                                       Daytona-Projected-Wet-AMD-1.jpg &     0.844429 \\
    11 &             AMD &                                                      \href{https://recognizingpathology.optos.com/wp-content/uploads/2018/02/Color-Wet-AMD-OD-California.jpg}{ORP} &                                                       Color-Wet-AMD-OD-California.jpg &     0.936324 \\
    12 &             AMD &                                                      \href{https://recognizingpathology.optos.com/wp-content/uploads/2018/02/Color-Wet-AMD-OS-California.jpg}{ORP} &                                                       Color-Wet-AMD-OS-California.jpg &     0.758085 \\
    13 &             AMD &                                          \href{https://recognizingpathology.optos.com/wp-content/uploads/2016/02/P200Tx-Projected-AMD-Geographic-Atrophy.jpg}{ORP} &                                           P200Tx-Projected-AMD-Geographic-Atrophy.jpg &     0.830430 \\
    14 &             AMD &                          \href{https://recognizingpathology.optos.com/wp-content/uploads/2016/04/AMD-Wet-California-Courtesy-Tim-Steffens-CRA-OCT-C-FOPS.jpg}{ORP} &                           AMD-Wet-California-Courtesy-Tim-Steffens-CRA-OCT-C-FOPS.jpg &     0.987892 \\
    15 &             AMD &                         \href{https://recognizingpathology.optos.com/wp-content/uploads/2016/02/Color-Atrophic-AMD-with-Geographic-Atrophy-OS-California.jpg}{ORP} &                          Color-Atrophic-AMD-with-Geographic-Atrophy-OS-California.jpg &     0.936971 \\
    16 &              DR &            \href{https://recognizingpathology.optos.com/wp-content/uploads/2016/02/P200Tx-Projected-Severe-NPDR-with-DME-HariprasadDiabeticRetinopathy22.jpg}{ORP} &             P200Tx-Projected-Severe-NPDR-with-DME-HariprasadDiabeticRetinopathy22.jpg &     0.869121 \\
    17 &              DR &                                      \href{https://recognizingpathology.optos.com/wp-content/uploads/2018/02/Color-NPDR-with-Macular-Edema-OD-California.jpg}{ORP} &                                       Color-NPDR-with-Macular-Edema-OD-California.jpg &     0.907351 \\
    18 &              DR &                       \href{https://recognizingpathology.optos.com/wp-content/uploads/2016/02/P200Tx-Projected-Diabetic-RetinopathyDiabeticRetinopathy16.jpg}{ORP} &                        P200Tx-Projected-Diabetic-RetinopathyDiabeticRetinopathy16.jpg &     0.886153 \\
    19 &              DR &                                      \href{https://recognizingpathology.optos.com/wp-content/uploads/2018/02/Color-NPDR-with-Macular-Edema-OS-California.jpg}{ORP} &                                       Color-NPDR-with-Macular-Edema-OS-California.jpg &     0.970982 \\
    20 &              DR &                            \href{https://recognizingpathology.optos.com/wp-content/uploads/2016/02/Daytona-Projected-PDR-with-PDP-1DiabeticRetinopathy11.jpg}{ORP} &                             Daytona-Projected-PDR-with-PDP-1DiabeticRetinopathy11.jpg &     0.847240 \\
    21 &              DR &                              \href{https://recognizingpathology.optos.com/wp-content/uploads/2016/02/P200Tx-Projected-DR-with-PRP-1DiabeticRetinopathy17.jpg}{ORP} &                               P200Tx-Projected-DR-with-PRP-1DiabeticRetinopathy17.jpg &     0.940622 \\
    22 &              DR &                                 \href{https://recognizingpathology.optos.com/wp-content/uploads/2016/02/California-DR-and-PRP-StangaDiabeticRetinopathy4.jpg}{ORP} &                                  California-DR-and-PRP-StangaDiabeticRetinopathy4.jpg &     0.928405 \\
    23 &              DR &                                                  \href{https://imagebank.asrs.org/file/27747/proliferative-diabetic-retinopathy-with-pre-retinal-hemorrhage}{ASRS} &                                                              ASRS-RIB-Image-27747.jpg &     0.955919 \\
    24 &              DR &                                                    \href{https://recognizingpathology.optos.com/wp-content/uploads/2018/02/Color-Severe-NPDR2-California.jpg}{ORP} &                                                     Color-Severe-NPDR2-California.jpg &     0.649301 \\
    25 &              DR &                \href{https://recognizingpathology.optos.com/wp-content/uploads/2016/02/California-Proliferative-Diabetic-RetinopathyDiabeticRetinopathy8.jpg}{ORP} &                 California-Proliferative-Diabetic-RetinopathyDiabeticRetinopathy8.jpg &     0.963178 \\
    26 &              DR &                                                         \href{https://recognizingpathology.optos.com/wp-content/uploads/2018/02/Color-PDR2-OS-California.jpg}{ORP} &                                                          Color-PDR2-OS-California.jpg &     0.942234 \\
    27 &              DR &                                                     \href{https://recognizingpathology.optos.com/wp-content/uploads/2018/02/Color-Severe-NPDR-California.jpg}{ORP} &                                                      Color-Severe-NPDR-California.jpg &     0.970263 \\
    28 &              DR &                                                         \href{https://recognizingpathology.optos.com/wp-content/uploads/2018/02/Color-PDR2-OD-California.jpg}{ORP} &                                                          Color-PDR2-OD-California.jpg &     0.933320 \\
    29 &              DR &                                       \href{https://recognizingpathology.optos.com/wp-content/uploads/2016/02/California-DR-Stanga-1DiabeticRetinopathy5.jpg}{ORP} &                                        California-DR-Stanga-1DiabeticRetinopathy5.jpg &     0.948500 \\
    30 &              DR &                                \href{https://recognizingpathology.optos.com/wp-content/uploads/2016/02/California-DR-and-PRP-Sadda-1DiabeticRetinopathy1.jpg}{ORP} &                                 California-DR-and-PRP-Sadda-1DiabeticRetinopathy1.jpg &     0.967951 \\
    31 &              DR &          \href{https://recognizingpathology.optos.com/wp-content/uploads/2016/02/P200Tx-Projected-Severe-NPDR-with-DME-Hariprasad-2DiabeticRetinopathy21.jpg}{ORP} &           P200Tx-Projected-Severe-NPDR-with-DME-Hariprasad-2DiabeticRetinopathy21.jpg &     0.954090 \\
    32 &            DR   &       \href{https://recognizingpathology.optos.com/wp-content/uploads/2016/02/P200Tx-Projected-Diabetic-Retinopathy-with-Mac-Grid-1DiabeticRetinopathy14.jpg}{ORP} &        P200Tx-Projected-Diabetic-Retinopathy-with-Mac-Grid-1DiabeticRetinopathy14.jpg &     0.949963 \\
    33 &          DR,AMD &              \href{https://recognizingpathology.optos.com/wp-content/uploads/2016/02/Color-Wet-AMD-and-NPDR-California-Courtesy-Mandar-Joshi-MD_result-1.jpg}{ORP} &              Color-Wet-AMD-and-NPDR-California-Courtesy-Mandar-Joshi-MD\_result-1.jpg &     0.968636 \\
    34 &          DR,AMD &           \href{https://recognizingpathology.optos.com/wp-content/uploads/2016/02/Color-Wet-AMD-and-NPDR-OS-California-Courtesy-Mandar-Joshi-MD_result-1.jpg}{ORP} &           Color-Wet-AMD-and-NPDR-OS-California-Courtesy-Mandar-Joshi-MD\_result-1.jpg &     0.945011 \\
    35 & DR,Drusen(AMD?) &                      \href{https://recognizingpathology.optos.com/wp-content/uploads/2016/02/Daytona-Projected-DR-Peripheral-DrusenDiabeticRetinopathy10.jpg}{ORP} &                       Daytona-Projected-DR-Peripheral-DrusenDiabeticRetinopathy10.jpg &     0.961227 \\
    36 &             Gla & \href{https://recognizingpathology.optos.com/wp-content/uploads/2018/02/Glaucoma-Suspect-California-William-Lesko-MD-North-Jersey-Eye-Associat..._result.jpg}{ORP} & Glaucoma-Suspect-California-William-Lesko-MD-North-Jersey-Eye-Associat...\_result.jpg &     0.823686 \\
    37 &         Healthy &                       \href{https://recognizingpathology.optos.com/wp-content/uploads/2017/01/Color-Healthy-California-Courtesy-Michael-Singer-MD_result.jpg}{ORP} &                       Color-Healthy-California-Courtesy-Michael-Singer-MD\_result.jpg &     0.844392 \\
    38 &         Healthy &                                               \href{https://recognizingpathology.optos.com/wp-content/uploads/2017/01/Color-Stereo-Healthy-OS-California.jpg}{ORP} &                                     Color-Stereo-Healthy-OS-California\_leftpanel.jpg &     0.727054 \\
    39 &         Healthy &                                                    \href{https://recognizingpathology.optos.com/wp-content/uploads/2017/01/California-optomap-am-Healthy.jpg}{ORP} &                                                     California-optomap-am-Healthy.jpg &     0.747208 \\
    40 &         Healthy &                                                                          \href{https://thefamilyvisioncenter.com/wp-content/uploads/2020/06/Technology-1.jpg}{Web} &                                                                      Technology-1.jpg &     0.610550 \\
    41 &         Healthy &                                                                           \href{https://www.optiqueofdenver.com/wp-content/uploads/2013/05/Color-Fundus1.jpg}{Web} &                                                                     Color-Fundus1.jpg &     0.445265 \\
    42 &         Healthy &                                               \href{https://recognizingpathology.optos.com/wp-content/uploads/2017/01/Color-Stereo-Healthy-OD-California.jpg}{ORP} &                                     Color-Stereo-Healthy-OD-California\_leftpanel.jpg &     0.769579 \\
    43 &         Healthy &                                                                                                                   \href{https://www.retinarocks.org/}{RetinaRocks} &                                                           Normal HOZ-20190131 (1).jpg &     0.578932 \\
    44 &         Healthy &                                           \href{https://recognizingpathology.optos.com/wp-content/uploads/2016/02/Daytona-Projected-Healthy-Retina-Adult.jpg}{ORP} &                                            Daytona-Projected-Healthy-Retina-Adult.jpg &     0.667379 \\
    45 &         Healthy &                                                             \href{https://recognizingpathology.optos.com/wp-content/uploads/2016/02/Color-Healthy-P200Tx.jpg}{ORP} &                                                              Color-Healthy-P200Tx.jpg &     0.494873 \\
    46 &         Healthy &                                                       \href{https://recognizingpathology.optos.com/wp-content/uploads/2016/02/Color-Healthy-Child-P200Tx.jpg}{ORP} &                                                        Color-Healthy-Child-P200Tx.jpg &     0.716572 \\
    47 &         Healthy &                                                         \href{https://recognizingpathology.optos.com/wp-content/uploads/2016/02/Color-Healthy-California.jpg}{ORP} &                                                          Color-Healthy-California.jpg &     0.581765 \\
    48 &         Healthy &                                                                                                  \href{https://imagebank.asrs.org/file/78895/healthy-retina}{ASRS} &                                                              ASRS-RIB-Image-78895.jpg &     0.684483 \\
    49 &              RD &                                                                      \href{https://imagebank.asrs.org/file/30011/optos-giant-tear-within-retinal-detachment}{ASRS} &                                                              ASRS-RIB-Image-30011.jpg &     0.920010 \\
    50 &              RD &                                                                                                                   \href{https://www.retinarocks.org/}{RetinaRocks} &                                                                  RRD NVG-20191210.jpg &     0.939834 \\
    51 &              RD &                                                                                                                   \href{https://www.retinarocks.org/}{RetinaRocks} &                                                              RRD GSV-20190213 (1).jpg &     0.922635 \\
    52 &              RD &                                                                                                                   \href{https://www.retinarocks.org/}{RetinaRocks} &               RRD WVI-20191105 (1) Recurrent with PVR and new small temporal hole.jpg &     0.945073 \\
    53 &              RD &                                                                                      \href{https://imagebank.asrs.org/file/26484/chronic-retinal-detachment}{ASRS} &                                                              ASRS-RIB-Image-26484.jpg &     0.935745 \\
    54 &              RD &                                                                                              \href{https://imagebank.asrs.org/file/25718/retinal-detachment}{ASRS} &                                                              ASRS-RIB-Image-25718.jpg &     0.940452 \\
    55 &              RD &                                                                                                                   \href{https://www.retinarocks.org/}{RetinaRocks} &                                                       RRD SLO-20131219 Giant tear.jpg &     0.901568 \\
    56 &              RP &                                                                                                                   \href{https://www.retinarocks.org/}{RetinaRocks} &                                                   RP LYV2-20181121 (5) \_ LEFTEYE.jpg &     0.974408 \\
    57 &              RP &                                                                                                                   \href{https://www.retinarocks.org/}{RetinaRocks} &                                                  RP LYV2-20181121 (5) \_ RIGHTEYE.jpg &     0.958171 \\
    58 &              RP &                                                                                                                   \href{https://www.retinarocks.org/}{RetinaRocks} &                                                               RP DVY-20190416 (2).jpg &     0.970836 \\
    59 &              RP &                                                                         \href{https://imagebank.asrs.org/file/18045/autosomal-dominant-retinitis-pigmentosa}{ASRS} &                                                              ASRS-RIB-Image-18045.jpg &     0.964437 \\
    60 &              RP &                                                                                                                 \href{https://i.redd.it/hij0f9pkqn441.jpg}{Reddit} &                                                                     hij0f9pkqn441.jpg &     0.952194 \\
    61 &              RP &                                                                 \href{https://imagebank.asrs.org/file/28324/retinitis-pigmentosa-with-cystoid-macular-edema}{ASRS} &                                                              ASRS-RIB-Image-28324.jpg &     0.928548 \\
    62 &              RP &                                                                                                                   \href{https://www.retinarocks.org/}{RetinaRocks} &             RP GSL1-20190731 (1) 20-30 OU With OCT and Optos with FAF \_ RIGHTEYE.jpg &     0.973185 \\
    63 &              RP &                                                                                                                   \href{https://www.retinarocks.org/}{RetinaRocks} &                                                      RP DVY-20190416 (1) From SCO.jpg &     0.958874 \\
    64 &              RP &                                                                                            \href{https://imagebank.asrs.org/file/27209/retinitis-pigmentosa}{ASRS} &                                                              ASRS-RIB-Image-27209.jpg &     0.912519 \\
    65 &              RP &                                                                                                                   \href{https://www.retinarocks.org/}{RetinaRocks} &              RP GSL1-20190731 (1) 20-30 OU With OCT and Optos with FAF \_ LEFTEYE.jpg &     0.970093 \\
    66 &             RVO &        \href{https://recognizingpathology.optos.com/wp-content/uploads/2016/02/P200Tx-Projected-Retinal-Vein-Occlusion-with-Ozurdex-injectionOcclusion31.jpg}{ORP} &         P200Tx-Projected-Retinal-Vein-Occlusion-with-Ozurdex-injectionOcclusion31.jpg &     0.816859 \\
    67 &             RVO &                                                                                                                   \href{https://www.retinarocks.org/}{RetinaRocks} &                                              BRVO Major URI-20200106 Extramacular.jpg &     0.932180 \\
    68 &             RVO &                                                                                                                   \href{https://www.retinarocks.org/}{RetinaRocks} &                         Unknown DSV-20190618 (1) 74YOF Possible multifocal BRVO's.jpg &     0.930758 \\
    69 &             RVO &                                                                                                                   \href{https://www.retinarocks.org/}{RetinaRocks} &  CRVO Ischemic PVM-20190709 (1) HM OD 20-50 OS Fresh CRVO OD Old resolved CRVO OS.jpg &     0.925773 \\
    70 &             RVO &                               \href{https://recognizingpathology.optos.com/wp-content/uploads/2016/02/P200Tx-Projected-Retinal-Vein-OcclusionOcclusion33.jpg}{ORP} &                                P200Tx-Projected-Retinal-Vein-OcclusionOcclusion33.jpg &     0.947042 \\
    71 &             RVO &                                                                                      \href{https://imagebank.asrs.org/file/66298/crvo-with-papillophlebitis}{ASRS} &                                                              ASRS-RIB-Image-66298.jpg &     0.946579 \\
    72 &             RVO &                                                            \href{https://recognizingpathology.optos.com/wp-content/uploads/2016/02/Color-BRVO-California.jpg}{ORP} &                                                             Color-BRVO-California.jpg &     0.960900 \\
    73 &             RVO &                                     \href{https://recognizingpathology.optos.com/wp-content/uploads/2016/02/Color-Hemi-Retinal-Vein-Occlusion-California.jpg}{ORP} &                                      Color-Hemi-Retinal-Vein-Occlusion-California.jpg &     0.944914 \\
    74 &             RVO &                                                                                                                   \href{https://www.retinarocks.org/}{RetinaRocks} &                                                           BRVO Major SZB-20191203.jpg &     0.929029 \\
    75 &             RVO &                                                                                                            \href{https://imagebank.asrs.org/file/25606/crvo}{ASRS} &                                                              ASRS-RIB-Image-25606.jpg &     0.889695 \\
\bottomrule
\end{tabular}

}
\end{adjustbox}
\end{table}%

\FloatBarrier
\subsection{Model generalises to unseen disease (held-out images showing Artery Occlusion)}
\label{sec:sup_AO_results}

We evaluated our model on the 21 images showing Atery Occlusion (AO). We had excluded them as including a label for a disease with only 21 available images would have been unlikely to be useful, particular when using a three-way data split on the patient-level. However, this allows us now to evaluate the model on an ``unseen'' disease. 11 of those 21 images show only AO, 10 also show another disease. As our model might just recognise the other diseases, we stratify our analysis accordingly. 

\begin{figure}[ht]
\centering
\includegraphics[width=0.8\textwidth]{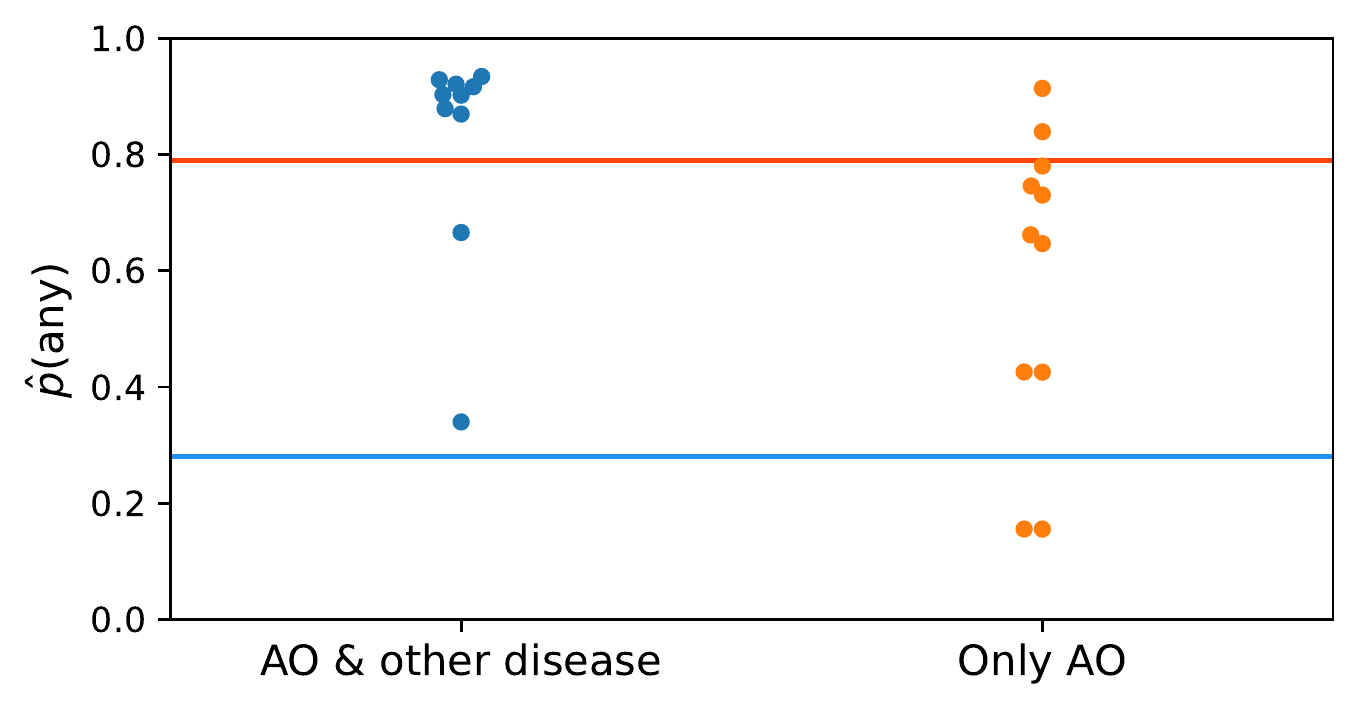}
\caption{Results of evaluating our model on the 21 held-out AO images showing the predicted probability of being diseased $\hat{p}(\mathrm{diseased})$ stratified by whether the images had another condition apart from AO. The red and blue horizontal lines plot indicate the conservative threshold $\hat{p}^t_\mathrm{conservative} =0.79$ and and less conserviative threshold $\hat{p}^t_\mathrm{less\kern 0.2em conservative} =0.28$, respectively.}
\label{fig:appendix_AOimages}
\end{figure}

We find that of 11 images that only show Artery Occlusion, our model correctly identifies 9 as diseased at the less conservative threshold but only 2 at the conservative threshold. The remaining 10 images show Artery Occlusion and a further disease, and our model identifies all 10 as diseased at the less conservative threshold, and 8 at the conservative incidence threshold. This is encouraging performance and highlights the value of the label for ``diseased'' as for some images, our model was not confident in any of the seven diseases that it was trained on, but nontheless very confident that there is some disease present. The generalisability to unseen diseases would be a very valuable property for practical applications, provided that the model does not frequently falsely flag up healthy patients.

\FloatBarrier
\subsection{Performance on subset of data used in a previous study on RP}
\label{sec:sup_RP_subset_previous_work}
We briefly experimented with a subset of 223 healthy controls and 150 images showing RP that were used in prior work on the TOP dataset \citep{masumoto2019accuracy}. We split the into train, validation and test sets containing 70, 15 and 15\% of the images. We find that even when using a ResNet18 with a linear prediction head, and only simple flip augmentations as regularisation, we can easily achieve perfect separation between the two classes on both the validation and test set with very high confidence in the correct labels (implying very low Brier scores). This matches what has been reported in the literature \citep{Antakibjophthalmol-2021-319030}. This perfect separation (AUC=1), achieved easily, with little tuning or effort in designing the model, contrasts with the very high yet not perfect separation our model obtained in our work (AUC=0.9438). We see this as an indication that our methodological improvements lead to a more realistic estimate of model performance.

\begin{figure}[!t]
     \centering
     \begin{subfigure}[b]{0.45\textwidth}
         \centering
         \includegraphics[width=\textwidth]{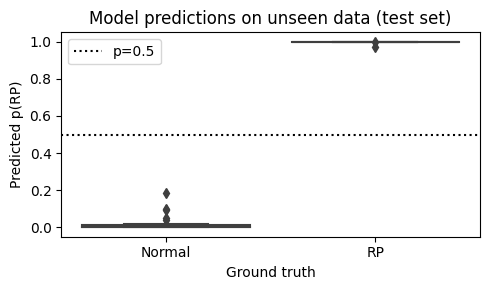}
         \caption{Model performance on the test set.}
     \end{subfigure}
     \hfill
          \begin{subfigure}[b]{0.45\textwidth}
         \centering
         \includegraphics[width=\textwidth]{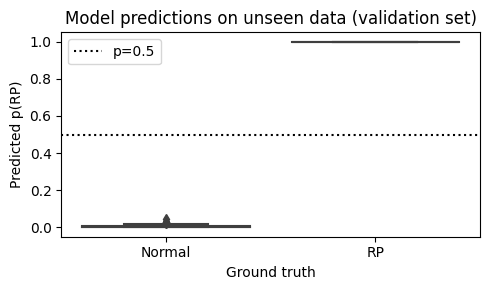}
         \caption{Model performance on the validation set.}
     \end{subfigure}
        \caption{Results of evaluating a simpler, binary model on a clean subset of RP images that was used in prior work \citep{masumoto2019accuracy}.}
        \label{fig:appendix_RP_subset_priorwork_performance}
\end{figure}

\FloatBarrier
\subsection{Assessment of Top 20 False Positives}
\label{sec:sup_TOP20FPs_assessment}

\begin{table}[!ht]%
\caption{Results of the assessment of the top 20 false positives. The scores indicates the grader's assessment whether the image seems to show pathology according to the following scale: -2: Definitely shows no pathology -1: Probably shows no pathology 0: Unclear 1: Probably shows pathology 2: Definitely shows pathology}
\label{tab:TOP20FP_assessment}
\centering
\begin{adjustbox}{max width=\textwidth, max totalheight=1\textheight-2\baselineskip}
{\small
\begin{tabular}{@{}llrrrr@{}}
\toprule
False positive & Filename & Score AM & Score IM & Score EP & Median score \\
\midrule
1                     & 003627\_01.jpg & 2   & 2   & 2    & 2    \\
2                     & 003626\_05.jpg & 2   & 2   & 2    & 2    \\
3                     & 002946\_02.jpg & 2   & 2   & 2    & 2    \\
4                     & 002947\_00.jpg & 1   & 2   & 2    & 2    \\
5                     & 000053\_01.jpg & 2   & 2   & 2    & 2    \\
6                     & 000440\_00.jpg & 0   & 1   & 2    & 1    \\
7                     & 003627\_00.jpg & 2   & 2   & 2    & 2    \\
8                     & 005008\_02.jpg & 2   & 2   & 1    & 2    \\
9                     & 000054\_01.jpg & 1   & 0   & 1    & 1    \\
10                    & 003626\_04.jpg & 2   & 2   & 2    & 2    \\
11                    & 001984\_00.jpg & -1  & -1  & 0    & -1   \\
12                    & 000686\_01.jpg & 0   & -1  & 0    & 0    \\
13                    & 001389\_01.jpg & 0   & -1  & 1    & 0    \\
14                    & 002313\_00.jpg & 2   & 2   & 2    & 2    \\
15                    & 001778\_00.jpg & -2  & -1  & 1    & -1   \\
16                    & 001436\_00.jpg & 2   & 2   & 2    & 2    \\
17                    & 001389\_00.jpg & -1  & -1  & -1   & -1   \\
18                    & 003376\_01.jpg & 1   & 1   & 1    & 1    \\
19                    & 004733\_00.jpg & 1   & 2   & 1    & 1    \\
20                    & 001566\_00.jpg & 2   & -1  & 0    & 0    \\
\midrule
Median                &                & 1.5 & 2   & 1.5  & 1.5  \\
Mean                  &                & 1   & 0.9 & 1.25 & 1.05 \\
Count (score\textgreater{0})&                & 14  & 13  & 16   & 14   \\ 
\bottomrule
\end{tabular}
}
\end{adjustbox}
\end{table}%

We assessed the top 20 most confident false positives our model generated on the test set, i.e. we selected the 20 images with the highest $\hat{p}(diseased)$ where the image had no disease according to the labels. This allows us to get a sense of the label noise in the TOP dataset.

The three co-authors with relevant experience each assessed these images independently. A.M. and I.M. are experienced clinicians and researchers in ophthalmology. E.P. is a researcher in retinal image analysis and completed her PhD on automated analysis of UWF images.  

The instructions for the assessment were as follows: 
\textit{"Please briefly evaluate whether each of the images seems to show pathology and assign a score from -2 to 2 according to the following scale. -2: Definitely shows no pathology -1: Probably shows no pathology 0: Unclear 1: Probably shows pathology 2: Definitely shows pathology"}

In \cref{sec:sup_TOP20FPs_assessment}, we report the results of this assessment. For 14 of the images, the median score was greater than 0, indicating that a majority of the graders thought that the image might show pathology. This suggests that these images are likely to show pathology. For 17 of the images, at least one of the graders thought that the image might show pathology. This assessment is limited as we only examined a small number of images. Furthermore, they were assessed by co-authors of this work. Despite our efforts to remain impartial, this could have introduced bias into the assessment.

\FloatBarrier
\subsection{Age+Sex benchmark model algorithm and hyperparameter selection}
\label{sec:sup_baselinehyperparams}
For Logistic Regression, we min-max scale input features to $[0,1]$ and choose the penalty type from $\{L1, L2, \mathrm{ElasticNet}\}$ and inverse penalty strength $C$ from $\{10, 1, 0.1\}$. ElasticNet is a combination of L1 (LASSO) and L2 (Ridge) penalties, which we weigh equally. For RFC, we grow 100 trees and choose number of features per tree from $\{\sqrt{n_\mathrm{features}}, 0.1 n_\mathrm{features}, n_\mathrm{features}\}$ and maximum tree depth from  $\{6, 10, \mathrm{Unlimited}\}$. For KNN, we choose the number of neighbours to consider $k$ from $\{5, 15, 30, 60\}$ and the distance measure used from $\{\mathrm{Manhattan}, \mathrm{Euclidean}\}$. We considered all classifiers and hyperparameter settings and selected the combination that performed best in terms of AUC on distinguishing unhealthy from healthy images on the validation set, the same criterion used for selecting the final DL model.

\end{document}